# Stabilization of uncertain linear distributed delay systems with dissipativity constraints*

Qian Feng, Sing Kiong Nguang

*Department of Electrical and Computer Engineering, The University of Auckland, Auckland 1010, New Zealand*

## Abstract

This paper examines the problem of stabilizing linear distributed delay systems with nonlinear distributed delay kernels and dissipativity constraints. Specifically, the nonlinear distributed kernel includes functions such as polynomials, trigonometric and exponential functions. By constructing a Liapunov-Krasovskiĭ functional related to the distributed kernels, sufficient conditions for the existence of a state feedback controller which stabilizes the uncertain distributed delay systems with dissipativity constraints are given in terms of linear matrix inequalities (LMIs). In contrast to existing methods, the proposed scenario is less conservative or requires a smaller number of decision variables based on the application of a newly derived integral inequality. Finally, numerical examples are presented to demonstrate the validity and effectiveness of the proposed methodology.

## 1. Introduction

Among many models of time delay systems (TDS) [16], distributed delay systems (DDS) cover a wide range of real time applications [23, 25]. For a rigorous treatment and benchmark results on the frequency domain approaches for linear TDS or DDS, see the monograph [22] and the references therein. With regards to time domain approaches, the construction of Liapunov-Krasovskiĭ functional [15] has been adopted as the most common method to undertake both stability analysis and controller synthesis. In particular, the Complete Liapunov Krasovskiĭ Functional (CKLF) [15, 18], which provides both sufficient and necessary stability conditions for a delay system such as $\dot{\boldsymbol{x}}(t) = A_1\boldsymbol{x}(t) + A_2\boldsymbol{x}(t-r), r \geq 0$, incorporates most of the existing proposed functionals as special cases. For a thorough treatise on the fundamental theories of CKLF and its mathematical derivation, see [18] and references therein.

In contrast to constructing a quadratic function within the context of semi-definite programming, the decision variables of CLKF possess infinite dimension which gives rise to significant difficulties to produce numerical results. In addition, the similar problems have been encountered in dealing with the non-constant distributed delay terms. By assuming constant decision variables with constant distributed delay terms in particular, finite dimension constraints can be automatically obtained which leads to conventional stability conditions denoted by

*Email addresses:* qfen204@aucklanduni.ac.nz (Qian Feng), sk.nguang@auckland.ac.nz (Sing Kiong Nguang)

LMIs. There has been a significant series of literature on this direction to perform either stability analysis or controller synthesis for linear DDS [9, 5, 7]. For a collection of the previous works on this topic, see the monographs [11, 4]. On the other hand, the results in [38] have demonstrated that certain linear DDS can be transformed into a corresponding system with only discrete delays. However, the aforementioned method inherits obvious conservatisms due to the presence of additional dynamics required by adding new states.

An alternative synthesis approach, predicated on the discretization scheme proposed in [14], is presented in [10] considering linear DDS with a piecewise constant distributed delay term. Moreover, by using the application of the full-block S-procedure [29], a novel synthesis result is presented in [26] to tackle systems with rational distributed delay kernels which are capable of dealing with general distributed terms via approximations. However, the derived stabilization conditions require $(A, B)$ to be controllable in [26] ($A$ is the delay free state space matrix and $B$ is the input gain matrix), thereby ensuring the induced conservatisms cannot be ignored. Finally, a systematic way to construct controllers for linear delay systems, having forwarding or backstepping structures, has been investigated in [17].

In this paper, we propose a method for stabilizing Linear DDS with nonlinear distributed kernels, which is achieved by constructing a standard complete Liapunov-Krasovskiĭ functional. In addition, a quadratic supply function [30, 4] and uncertainty with full block constraints [31] having non-conservative assumptions are incorporated to provide a broad characterization of both controller objectives and robustness. Furthermore, a new integral inequality is derived for the formulation of the synthesis conditions, which can be considered as a generalization of the recently proposed Bessel-Legendre inequality [32, 33]. By applying the new derived integral inequality with the Projection Lemma [12], convex synthesis conditions can be derived in terms of LMIs. Unlike existing methods, the proposed solution neither requires $(A, B)$ to be controllable as in [26], nor demands forwarding or backstepping structures as in [17]. Furthermore, it can produce feasible results without considering approximations and with fewer decision variables in comparison with the stability analysis results in [35].

The paper is organized as follows: Some important preliminaries and the formulations of the synthesis problem are presented in Section 2. Section 3 contains the main results relating to controller synthesis. To demonstrate the validity and effectiveness of the proposed methodologies, numerical examples are investigated in Section 4 before the final conclusion in Section 5.

**Notation :** The notations in this paper follow standard rules, though certain new symbols will be introduced for the sake of compactness : $\mathbb{T} := \{x \in \mathbb{R} : x \geq 0\}$; $\mathbb{S}^n := \{X \in \mathbb{R}^{n\times n} : X = X^\top\}$; $\mathbb{R}^{n\times n}_{[n]} := \{X \in \mathbb{R}^{n\times n} : \operatorname{rank}(X) = n\}$; notations $\|\mathbf{x}\|_q = (\sum_{i=1}^n |x_i|^q)^{\frac{1}{q}}$ and $\|f(\cdot)\|_p = \left(\int_{\mathbb{R}} |f(t)|^p \mathsf{d}t\right)^{\frac{1}{p}}$ and $\|\boldsymbol{f}(\cdot)\|_{p|q} = \left(\int_{\mathbb{R}} \|\boldsymbol{f}(t)\|_q^p \mathsf{d}t\right)^{\frac{1}{p}}$ are the norms associated with $\mathbb{R}^n$ and Lebesgue integrable functions spaces $\mathcal{L}_p(\mathbb{R};\mathbb{R})$ and $\mathcal{L}_p(\mathbb{R};\mathbb{R}^n_{\|q})$, respectively. $\mathcal{C}(\mathcal{X};\mathbb{R}^n)$ with $\sup_{\tau\in\mathcal{X}} \|\boldsymbol{f}(\tau)\|_2$ is the Banach space of continuous functions with an uniform norm. $\mathbf{Sy}(X) := X + X^\top$ is the sum of a matrix with its transpose. A column vector containing a sequence of objects is defined as $\mathbf{col}_{i=1}^n x_i := \left[\mathbf{row}_{i=1}^n x_i^\top\right]^\top = \left[x_1^\top \cdots x_i^\top \cdots x_n^\top\right]^\top$. We use $*$ to denote $[*]YX = X^\top Y X$ or $X^\top Y[*] = X^\top Y X$. The direct sum of two matrices and $n$ matrices are defined as $X \oplus Y = \mathsf{Diag}(X, Y)$, $\bigoplus_{i=1}^n X_i = \mathsf{Diag}_{i=1}^n(X_i)$, respectively. Finally, $\otimes$ indicates the Kronecker product.

## 2. Prelimilaries and Problem Formulations

Without losing generality, we only consider a system with one delay channel in this paper for the sake of simplicity. However, one can easily extend the corresponding Krasovskiĭ functional to handle multiple delay channels simultaneously.

Consider a linear model of DDS

$$\begin{aligned}\dot{\boldsymbol{x}}(t) &= A_1\boldsymbol{x}(t) + A_2\boldsymbol{x}(t-r) + \int_{-r}^{0} \widetilde{A}_3(\tau)\boldsymbol{x}(t+\tau)\mathsf{d}\tau + B_1\boldsymbol{u}(t) + B_2\boldsymbol{w}(t)\\ \boldsymbol{z}(t) &= C_1\boldsymbol{x}(t) + C_2\boldsymbol{x}(t-r) + \int_{-r}^{0} \widetilde{C}_3(\tau)\boldsymbol{x}(t+\tau)\mathsf{d}\tau + D_1\boldsymbol{u}(t) + D_2\boldsymbol{w}(t)\\ \boldsymbol{x}(\tau) &= \boldsymbol{\phi}(\tau),\ \ \forall\tau \in \mathcal{O} := [-r, 0]_{\mathbb{R}}\end{aligned} \tag{1}$$

to be stabilized, where $\boldsymbol{x}(t) \in \mathbb{R}^n$ is the solution of (1), $\boldsymbol{u}(t) \in \mathbb{R}^p$ denotes input signals, $\boldsymbol{w}(\cdot) \in \mathcal{L}_2(\mathbb{T}; \mathbb{R}^q_{\|2})$ represents disturbance, $\boldsymbol{z}(t) \in \mathbb{R}^m$ is the regulated output, and $\boldsymbol{\phi}(\cdot) \in \mathcal{C}(\mathcal{O}; \mathbb{R}^n)$ denotes initial condition. Matrices $A_1; A_2; \widetilde{A}_3(\tau) \in \mathbb{R}^{n\times n}$, $C_1; C_2; \widetilde{C}_3(\tau) \in \mathbb{R}^{m\times n}$, $B_1 \in \mathbb{R}^{n\times p}$, $B_2 \in \mathbb{R}^{n\times q}$, $D_1 \in \mathbb{R}^{m\times p}$, $D_2 \in \mathbb{R}^{m\times q}$ are given state space systems parameters with $n; m; p; q \in \mathbb{N}$. $r \in \mathbb{T}$ is a given constant specifying the length of delay channel. Finally, $\widetilde{A}_3(\tau)$ and $\widetilde{C}_3(\tau)$ satisfy the following assumption.

**Assumption 1.** There exists $\boldsymbol{m}(\cdot) \in \mathcal{C}^{[1]}(\mathcal{O}; \mathbb{R}^\rho)$ with $\rho \in \mathbb{N}$ and $A_3 \in \mathbb{R}^{n\times n\rho}$, $C_3 \in \mathbb{R}^{m\times n\rho}$ such that $\forall\tau \in \mathcal{O}$, $\mathbb{R}^{n\times n} \ni \widetilde{A}_3(\tau) = A_3(\boldsymbol{m}(\tau) \otimes I_n)$ and $\mathbb{R}^{m\times n} \ni \widetilde{C}_3(\tau) = C_3(\boldsymbol{m}(\tau) \otimes I_n)$. In addition, $\boldsymbol{m}(\tau)$ satisfies the following properties:

$$\boldsymbol{m}(\tau) := \operatorname*{\mathbf{col}}_{i=1}^{\rho} m_i(\cdot), \quad \frac{\mathsf{d}\boldsymbol{m}(\tau)}{\mathsf{d}\tau} = \mathsf{M}\boldsymbol{m}(\tau), \tag{2}$$

where $\mathsf{M} \in \mathbb{R}^{\rho\times\rho}$ and $m_i(\tau)$ are linearly independent and that for all $i = 1\cdots\rho$ there exists an uncountable set $\mathcal{P} \subseteq \mathcal{O}$ such that $\forall\vartheta \in \mathcal{P},\ m_i(\vartheta) \neq 0$. Similar assumptions can be found in [34, 19, 24, 21].

**Remark 1.** Specifically, the elements inside of $\boldsymbol{m}(\tau)$ are the solutions of linear homogeneous differential equations with constant coefficients such as polynomials, trigonometric and exponential functions. In addition, there is no limitation on the size of the dimension of $\boldsymbol{m}(\tau)$ as long as it is able to cover all the elements in the distributed terms in (1). As for the generality of $\boldsymbol{m}(\tau)$, there are many applications that can be modeled by (1) compatible with Assumption 1, for example, the compartmental dynamic systems with distributed delays mentioned in [36]. Furthermore, distributed delay systems concerning gamma distributions in [26] and [13] with a finite delay range can be stabilized by the proposed methods as well.

In this paper, we assume that all states are available for feedback and (1) is stabilized by a state feedback controller $\boldsymbol{u}(t) = K\boldsymbol{x}(t)$ with $K \in \mathbb{R}^{p\times n}$. Substituting $\boldsymbol{u}(t) = K\boldsymbol{x}(t)$ into (1) and considering Assumption (1) yields

$$\begin{aligned}\dot{\boldsymbol{x}}(t) &= \Pi_1\boldsymbol{x}(t) + A_2\boldsymbol{x}(t-r) + \int_{-r}^{0} A_3M(\tau)\boldsymbol{x}(t+\tau)\mathsf{d}\tau + B_2\boldsymbol{w}(t)\\ \boldsymbol{z}(t) &= \Omega_1\boldsymbol{x}(t) + C_2\boldsymbol{x}(t-r) + \int_{-r}^{0} C_3M(\tau)\boldsymbol{x}(t+\tau)\mathsf{d}\tau + D_2\boldsymbol{w}(t)\end{aligned} \tag{3}$$

as the corresponding closed loop system, where $\Pi_1 = A_1 + B_1K$ and $\Omega_1 = C_1 + D_1K$.

To specify performance objectives for (3), we apply the quadratic form

$$\mathsf{s}(\boldsymbol{z}(t), \boldsymbol{w}(t)) = -\begin{bmatrix}\boldsymbol{z}(t)\\ \boldsymbol{w}(t)\end{bmatrix}^\top \mathbf{J}\begin{bmatrix}\boldsymbol{z}(t)\\ \boldsymbol{w}(t)\end{bmatrix}, \tag{4}$$

with

$$\mathbf{J} = \begin{bmatrix}J_1^{-1} & J_2\\ J_2^\top & J_3\end{bmatrix} \in \mathbb{S}^{(m+q)},\ \ J_1^{-1} \succeq 0 \tag{5}$$

considered in [30] to be the supply rate function.

The supply function in (4) is able to characterize numerous optimization constraints such as $\mathscr{L}_2$ gain control:

$$\sup_{\boldsymbol{w}(\cdot)\in\mathscr{L}_2(\mathbb{T};\mathbb{R}^q_{\|2})}\left(\frac{\|\boldsymbol{z}(\cdot)\tilde{\|}_{2|2}}{\|\boldsymbol{w}(\cdot)\tilde{\|}_{2|2}}\right)<\gamma \quad \text{with } \gamma>0 \text{ and } \quad J_1=\gamma I_m,\ J_3=-\gamma I_q,\ J_2=\mathsf{O}_{m\times q}; \tag{6}$$

and Sector Constraints when $m=q$ with

$$J_1^{-1}=I_m,\ J_2=-\frac{1}{2}(\alpha+\gamma)I_m,\ J_3=-\alpha\gamma I_m. \tag{7}$$

For the situation when $J_1^{-1}=J_3=\mathsf{O}_m$ and $J_2=-I_m$ with $m=q$, which corresponds to having the strict passivity constraint, the well posedness of $J_1$ does not need to be considered since there is no reason to apply Schur complement here given the fact that $\boldsymbol{z}^\top(t)J_1^{-1}\boldsymbol{z}(t)=0$. As a result, one can only use $J_1^{-1}=\mathsf{O}_m$ in (4) directly and no mathematical complications will be introduced in deriving the corresponding synthesis conditions.

The following lemmas and definition are required for the mathematical derivations in this paper.

**Lemma 1.** *$\forall P\in\mathbb{R}^{p\times q}$ and $\forall Q\in\mathbb{R}^{n\times m}$, we have*

$$(P\otimes I_n)(I_q\otimes Q)=(I_p\otimes Q)(P\otimes I_m). \tag{8}$$

*Moreover, we have $\forall X\in\mathbb{R}^{n\times m},\ \ \forall Y\in\mathbb{R}^{m\times p}$*

$$(XY)\otimes I_n=(X\otimes I_n)(Y\otimes I_n). \tag{9}$$

*Proof.* The aforementioned equalities are the special cases of the property of the Kronecker product: $(A\otimes B)(C\otimes D)=(AC)\otimes(BD)$. ■

**Lemma 2** (Projection Lemma)**.** *[12] Given $n;p;q\in\mathbb{N}$, $\Pi\in\mathbb{S}^n, P\in\mathbb{R}^{q\times n}, Q\in\mathbb{R}^{p\times n}$, there exists $\Theta\in\mathbb{R}^{p\times q}$ such that the following two propositions are equivalent :*

$$\Pi+P^\top\Theta^\top Q+Q^\top\Theta P\prec 0, \tag{10}$$

$$P_\perp^\top\Pi P_\perp\prec 0 \ \ \textit{and}\ \ Q_\perp^\top\Pi Q_\perp\prec 0, \tag{11}$$

*where $P_\perp$ and $Q_\perp$ are matrices in which the columns contain any basis of null space of matrix $P$ and $Q$, respectively.*

*Proof.* Refer to [12] and [4]. ■

To derive the synthesis conditions incorporating performance objectives, it requires the following stability criterion for functional differential equations [15] and the definition of dissipativity systems [39, 4].

**Lemma 3.** *Given $\mathcal{O}:=[-r,0]$ with $r>0$, a delay system $\dot{\boldsymbol{x}}(t)=\boldsymbol{f}(\boldsymbol{x}(t+\cdot))$, $\boldsymbol{x}(t+\cdot)\in\mathcal{C}(\mathcal{O};\mathbb{R}^n)$, $\forall t\in\mathbb{R}$ is uniformly globally asymptotically stable at its origin if there exists a differentiable functional $\mathsf{v}(\bullet):\mathcal{C}(\mathcal{O};\mathbb{R}^n)\to\mathbb{R}$ with $\mathsf{v}(\mathbf{0}(\cdot))=0$ such that the following conditions are satisfied:*

$$\exists\epsilon_1;\epsilon_2>0,\ \forall\boldsymbol{\phi}(\cdot)\in\mathcal{C}(\mathcal{O};\mathbb{R}^n):\epsilon_1\|\boldsymbol{\phi}(0)\|_2^2\le\mathsf{v}(\boldsymbol{\phi}(\cdot))\le\epsilon_2\sup_{\tau\in\mathcal{O}}\|\boldsymbol{\phi}(\tau)\|_2^2, \tag{12}$$

$$\exists\epsilon_3<0,\forall\boldsymbol{\phi}(\cdot)\in\mathcal{C}(\mathcal{O};\mathbb{R}^n):\dot{\mathsf{v}}(\boldsymbol{\phi}(\cdot))\le\epsilon_3\|\boldsymbol{\phi}(0)\|_2^2. \tag{13}$$

*Proof.* See [15] for the proof of the general Liapunov-Krasovskiĭ stability theorem which encompasses (12) and (13) as special cases. ■

**Definition 1.** [39, 4] Given $\mathcal{O} := [-r, 0]$ with $r > 0$, a delay system $\dot{\boldsymbol{x}}(t) = \boldsymbol{f}\big(\boldsymbol{x}(t+\cdot), \boldsymbol{w}(t)\big)$, $\boldsymbol{x}(t+\cdot) \in \mathcal{C}(\mathcal{O}; \mathbb{R}^n)$, $\forall t \in \mathbb{R}$ with a output $\boldsymbol{z}(t) = \boldsymbol{g}\big(\boldsymbol{x}(t+\cdot), \boldsymbol{w}(t)\big)$ is dissipative with respect to the supply function $\mathsf{s}(\boldsymbol{z}(t), \boldsymbol{w}(t))$ if there exists a differentiable functional $\mathsf{v}(\bullet) : \mathcal{C}(\mathcal{O}; \mathbb{R}^n) \to \mathbb{R}$ such that

$$\forall t \in \mathbb{T}, \quad \dot{\mathsf{v}}(\boldsymbol{x}(t+\cdot)) - \mathsf{s}(\boldsymbol{z}(t), \boldsymbol{w}(t)) \leq 0, \tag{14}$$

where (14) is an equivalent condition of the original definition of dissipativity, given $\mathsf{v}(\bullet) : \mathcal{C}(\mathcal{O}; \mathbb{R}^n) \to \mathbb{T}$ is differentiable.

**Lemma 4.** *Given a uncountable closed set $\mathcal{K} \subseteq \mathbb{R}$ and a function series $\{m_i(\tau)\}_{i=1}^d \in \mathcal{L}_2(\mathcal{K}; \mathbb{R})$, $d \in \mathbb{N}$ such that $m_i(\tau)$ are linearly independent and that $\forall i = 1 \cdots d$, there exists an uncountable set $\mathcal{P} \subseteq \mathcal{K}$ such that $\forall \vartheta \in \mathcal{P},\ m_i(\vartheta) \neq 0$, then the inequality*

$$\int_{\mathcal{K}} \boldsymbol{m}(\tau)\boldsymbol{m}^\top(\tau)\mathrm{d}\tau \succ 0 \tag{15}$$

*holds, where $\boldsymbol{m}(\tau) := \mathbf{col}_{i=1}^d\, m_i(\tau)$.*

*Proof.* We apply proof by contradiction. It is obvious that $\boldsymbol{m}(\tau)\boldsymbol{m}^\top(\tau) \succeq 0, \forall \tau \in \mathcal{K}$. As a result, we have $\int_{\mathcal{K}} \boldsymbol{m}(\tau)\boldsymbol{m}^\top(\tau)\mathrm{d}\tau \succeq 0$, based on the properties of quadratic form. Now we only consider the situation of semi-positive definite, this implies that $\exists \mathbf{p} \in \mathbb{R}^d \setminus \{\mathbf{0}\}$ such that $\int_{\mathcal{K}} \mathbf{p}^\top \boldsymbol{m}(\tau)\boldsymbol{m}^\top(\tau)\mathbf{p}\mathrm{d}\tau = \int_{\mathcal{K}} \left[\mathbf{p}^\top \boldsymbol{m}(\tau)\right]^2 \mathrm{d}\tau = 0$. Furthermore, it follows that $\int_{\mathcal{K}} \left[\mathbf{p}^\top \boldsymbol{m}(\tau)\right]^2 \mathrm{d}\tau = 0$ if and only if $\mathbf{p}^\top \boldsymbol{m}(\tau) = 0$ due to the fact that $m_i(\tau)$ have nonzero values on an uncountable set $\mathcal{P} \subseteq \mathcal{K}$. However, it is impossible to have $\mathbf{p}^\top \boldsymbol{m}(\tau) = 0$ since $m_i(\tau)$ are linearly independent and $\mathbf{p} \neq \mathbf{0}$. This is a contradiction which implies (15). ■

**Lemma 5.** *Given a uncountable closed set $\mathcal{K} \subseteq \mathbb{R}$ and $U \in \mathbb{S}^n_{\succeq 0}$ and a function series $\{m_i(\tau)\}_{i=1}^d \in \mathcal{L}_2(\mathcal{K}; \mathbb{R})$ with $d \in \mathbb{N}$ such that $m_i(\tau)$ are linearly independent and that $\forall i = 1 \cdots d$, there exists an uncountable set $\mathcal{P} \subseteq \mathcal{K}$ such that $\forall \vartheta \in \mathcal{P},\ m_i(\vartheta) \neq 0$, then $\forall \boldsymbol{x}(\cdot) \in \mathcal{L}_2(\mathcal{K}; \mathbb{R}^n_{\|2})$ we have*

$$\int_{\mathcal{K}} \boldsymbol{x}^\top(\tau)U\boldsymbol{x}(\tau)\mathrm{d}\tau \geq \int_{\mathcal{K}} \boldsymbol{x}^\top(\tau)M^\top(\tau)\mathrm{d}\tau\, (\mathsf{F} \otimes U) \int_{\mathcal{K}} M(\tau)\boldsymbol{x}(\tau)\mathrm{d}\tau, \tag{16}$$

*where $\mathsf{F}^{-1} = \int_{\mathcal{K}} \boldsymbol{m}(\tau)\boldsymbol{m}^\top(\tau)\mathrm{d}\tau \succ 0$ and $M(\tau) := \boldsymbol{m}(\tau) \otimes I_n$ such that $\boldsymbol{m}(\tau) := \left[\mathbf{col}_{i=1}^d\, m_i(\tau)\right]$.*

*Proof.* See Appendix A for details. It is important to mention that the vector $\boldsymbol{m}(\tau)$ in Assumption 1 satisfies $\int_{-r}^0 \boldsymbol{m}(\tau)\boldsymbol{m}^\top(\tau) \succ 0$ since it satisfies all the prerequisite conditions in Lemma 3 and $\mathcal{C}^{[1]}([-r, 0]; \mathbb{R}) \subset \mathcal{L}_2([-r, 0]; \mathbb{R})$. ■

**Remark 2.** Given $\mathcal{K} = [-r, 0], r > 0$ and $\boldsymbol{m}(\tau) := \mathbf{col}_{i=0}^\rho\, \ell_i(\tau)$ with the Legendre polynomials

$$\ell_d(\tau) = (-1)^d \sum_{k=0}^d (-1)^k \binom{d}{k}\binom{d+k}{k}\left(\frac{\tau+r}{r}\right)^k, \quad \forall d \in \mathbb{N} \cup \{0\}, \quad \forall \tau \in [-r, 0], \tag{17}$$

then the result in (16) holds with $\mathsf{F}^{-1} = \bigoplus_{i=0}^\rho \frac{r}{2i+1}$. This demonstrates the fact that (16) can be regarded as a generalization of the Bessel-Legendre Inequality proposed in [32, 33]. Indeed, if the function series $\{m_i(\tau)\}_{i=1}^d \in \mathcal{L}_2(\mathcal{K}; \mathbb{R})$ in Lemma 4 contains only orthogonal functions, then $\int_{\mathcal{K}} \boldsymbol{m}(\tau)\boldsymbol{m}^\top(\tau)\mathrm{d}\tau$ must be a diagonal matrix. Furthermore, (16) can also be considered as a generalization of the results in [28]. Finally, it is worthy to emphasize that a discrete version of (16)

$$\sum_{k \in \mathcal{J}} \boldsymbol{x}^\top(k)U\boldsymbol{x}(k) \geq [*]\, (\mathsf{F} \otimes U) \left[\sum_{k \in \mathcal{J}} M(k)\boldsymbol{x}(k)\right], \quad \mathsf{F}^{-1} = \sum_{k \in \mathcal{J}} \boldsymbol{m}(k)\boldsymbol{m}^\top(k), \quad \mathcal{J} \subseteq \mathbb{Z} \tag{18}$$

can be easily derived given the connections between integrations and summations.

For the sake of compactness, denoting $\widehat{\mathsf{M}} = \mathsf{M} \otimes I_n$ and applying (9) to (2), we have

$$\dot{M}(\tau) = \widehat{\mathsf{M}} M(\tau). \tag{19}$$

## 3. Main Results on Controller Synthesis

To prove global asymptotic stability at the origin for the closed loop system (3), we consider a parameterized functional of the form

$$\mathsf{v}(\boldsymbol{x}(t+\cdot)) := \left[*\right]\begin{bmatrix} P & Q \\ * & R \end{bmatrix}\begin{bmatrix} \boldsymbol{x}(t) \\ \int_{-r}^{0} M(\tau)\boldsymbol{x}(t+\tau)\mathsf{d}\tau \end{bmatrix} + \int_{-r}^{0} \boldsymbol{x}^\top(t+\tau)\Big[S+(\tau+r)U\Big]\boldsymbol{x}(t+\tau)\mathsf{d}\tau \tag{20}$$

to be constructed, where $P \in \mathbb{S}^n,\ Q \in \mathbb{R}^{n\times \rho n},\ R \in \mathbb{S}^{\rho n},\ S; U \in \mathbb{S}^n$.

By utilizing Lemma 3 and Definition 1 with (20) and (4), sufficient conditions for the existence of a state feedback controller which stabilizes the closed loop system (3) and takes into account the supply rate function in (4) are derived in terms of LMIs, which is elaborated in the following theorem.

**Theorem 1.** *Given parameters $J_1^{-1} \in \mathbb{S}^m_{\succ 0}$, $J_2 \in \mathbb{R}^{m\times q}$, $J_3 \in \mathbb{R}^{q\times q}$ in (4) and $\eta_1; \eta_2 \in \mathbb{R}$, $\{\epsilon_i\}_{i=1}^{\rho} \in \mathbb{R}$, $\rho \in \mathbb{N}$ and the open loop system in (1) satisfying all the properties mentioned in Assumption 1, the closed loop system in (3) is globally asymptotically stable at the origin with $\boldsymbol{w}(t) \equiv \mathbf{0}_q$, and dissipative with (4), if there exist $\check{P} \in \mathbb{S}^n$, $X \in \mathbb{R}^{n\times n}$, $V \in \mathbb{R}^{p\times n}$, $\check{Q} \in \mathbb{R}^{n\times \rho n}$, $\check{R} \in \mathbb{S}^{\rho n}$, $\check{S}; \check{U} \in \mathbb{S}^n$ such that*

$$\begin{bmatrix} \check{P} & \check{Q} \\ * & \check{R} + \mathsf{F}\otimes \check{S} \end{bmatrix} \succ 0,\ \ \check{S} \succ 0,\ \ \check{U} \succ 0, \tag{21}$$

$$\boldsymbol{\Theta} := \mathbf{Sy}\left[\widetilde{\mathbf{I}}^\top \widetilde{\mathbf{Y}}\right] + \begin{bmatrix} \mathsf{O}_n & \widetilde{\mathbf{P}} \\ * & \widetilde{\boldsymbol{\Phi}} \end{bmatrix} \prec 0, \tag{22}$$

*where* $\mathsf{F}^{-1} := \int_{-r}^{0} \boldsymbol{m}(\tau)\boldsymbol{m}^\top(\tau)\mathsf{d}\tau$ *and*

$$\begin{aligned}
&\widetilde{\mathbf{I}} := \Big[I_n \quad \eta_1 I_n \quad \eta_2 I_n \quad \mathop{\mathbf{row}}_{i=1}^{\rho} \epsilon_i I_n \quad \mathsf{O}_{n\times q} \quad \mathsf{O}_{n\times m}\Big],\ \ \widetilde{\mathbf{P}} := \Big[\check{P} \quad \mathsf{O}_n \quad \check{Q} \quad \mathsf{O}_{n\times q} \quad \mathsf{O}_{n\times m}\Big],\\
&\widetilde{\mathbf{Y}} := \Big[-X \quad A_1X + B_1V \quad A_2X \quad A_3(I_\rho\otimes X) \quad B_2 \quad \mathsf{O}_{n\times m}\Big],\\
&\widetilde{\boldsymbol{\Phi}} := \mathbf{Sy}\left(\begin{bmatrix} \check{Q} \\ \mathsf{O}_{n\times n\rho} \\ \check{R} \\ \mathsf{O}_{(q+m)\times n\rho} \end{bmatrix}\Big[M(0) \quad -M(-r) \quad -\widehat{\mathsf{M}} \quad \mathsf{O}_{n\rho\times(q+m)}\Big]\right)\\
&\quad + \Big[(\check{S}+r\check{U})\Big] \oplus \Big[-\check{S}\Big] \oplus \Big[-\mathsf{F}\otimes\check{U}\Big] \oplus J_3 \oplus (-J_1)\\
&\quad + \mathbf{Sy}\left(\begin{bmatrix} \mathsf{O}_{(2n+n\rho)\times m} \\ J_2^\top \\ I_m \end{bmatrix}\Big[C_1X + D_1V \quad C_2X \quad C_3(I_\rho\otimes X) \quad D_2 \quad \mathsf{O}_m\Big]\right).
\end{aligned} \tag{23}$$

*Furthermore the controller parameter is calculated via $K = VX^{-1}$.*

*Proof.* The subsequent proof is based on constructing $\mathsf{v}(\boldsymbol{x}(t+\cdot))$ in (20) incorporating $\mathsf{s}(\boldsymbol{z}(t), \boldsymbol{w}(t))$ in (4), which consists of two major parts. We will first prove that (22) implies both (13) and (14). Subsequently, we prove that (21) implies (12) in *Part 2*. The existence of the upper bound of $\mathsf{v}(\boldsymbol{x}(t+\cdot)$ can be independently proved without considering the synthesis conditions in Theorem 1.

*Part 1.*

First we derive the conditions implying both (13) and (14). To begin with, reformulate $\mathsf{s}(\boldsymbol{z}(t), \boldsymbol{w}(t))$ in (4) into

$$-\mathsf{s}(\boldsymbol{z}(t), \boldsymbol{w}(t)) = \boldsymbol{z}^\top(t)J_1^{-1}\boldsymbol{z}(t) + \mathbf{Sy}\left[\boldsymbol{z}^\top(t)J_2\boldsymbol{w}(t)\right] + \boldsymbol{w}^\top(t)J_3\boldsymbol{w}(t). \tag{24}$$

Clearly, only term $\boldsymbol{z}^\top(t)J_1^{-1}\boldsymbol{z}(t)$ in (24) introduces nonlinearity with respect to the undetermined variables in (3) and (20). Substituting $\boldsymbol{z}(t)$ in equation (3) into $\boldsymbol{z}^\top(t)J_1^{-1}\boldsymbol{z}(t)$ produces

$$\boldsymbol{z}^\top(t)J_1^{-1}\boldsymbol{z}(t) = \boldsymbol{\chi}^\top(t)\,[*]\,J_1^{-1}\begin{bmatrix}\Omega_1 & C_2 & C_3 & D_2\end{bmatrix}\boldsymbol{\chi}(t), \tag{25}$$

where

$$\boldsymbol{\chi}(t) := \mathbf{col}\left[\boldsymbol{x}(t), \boldsymbol{x}(t-r), \int_{-r}^{0}\boldsymbol{\varphi}(t+\tau)\mathsf{d}\tau, \boldsymbol{w}(t)\right], \quad \boldsymbol{\varphi}(t+\tau) = M(\tau)\boldsymbol{x}(t+\tau). \tag{26}$$

Now differentiate $\mathsf{v}(\boldsymbol{x}(t+\cdot))$ alongside the trajectory of the closed loop system (3) and consider both Assumption 1 and (24) with the relations (25) and

$$\frac{\mathsf{d}}{\mathsf{d}t}\int_{-r}^{0} M(\tau)\boldsymbol{x}(t+\tau)\mathsf{d}\tau = M(0)\boldsymbol{x}(t) - M(-r)\boldsymbol{x}(t-r) - \widehat{\mathsf{M}}\int_{-r}^{0} M(\tau)\boldsymbol{x}(t+\tau)\mathsf{d}\tau. \tag{27}$$

Then it yields

$$\begin{aligned}
&\dot{\mathsf{v}}(\boldsymbol{x}(t+\cdot)) - \mathsf{s}(\boldsymbol{z}(t), \boldsymbol{w}(t)) \\
&= \boldsymbol{\chi}^\top(t)\,\mathbf{Sy}\left(\begin{bmatrix} I_n & \mathsf{O}_{n\times n\rho} \\ \mathsf{O}_n & \mathsf{O}_{n\times n\rho} \\ \mathsf{O}_{n\rho\times n} & I_{n\rho} \\ \mathsf{O}_{q\times n} & \mathsf{O}_{q\times n\rho}\end{bmatrix}\begin{bmatrix} P & Q \\ * & R\end{bmatrix}\begin{bmatrix}\Pi_1 & A_2 & A_3 & B_2 \\ M(0) & -M(-r) & -\widehat{\mathsf{M}} & \mathsf{O}_{n\rho\times q}\end{bmatrix}\right)\boldsymbol{\chi}(t) \\
&+ \boldsymbol{x}^\top(t)\,(S + rU)\,\boldsymbol{x}(t) - \boldsymbol{x}^\top(t-r)S\boldsymbol{x}(t-r) - \int_{-r}^{0}\boldsymbol{x}^\top(t+\tau)U\boldsymbol{x}(t+\tau)\mathsf{d}\tau + (25) \\
&+ \boldsymbol{\chi}^\top(t)\left[*\right]\begin{bmatrix}\mathsf{O}_m & J_2 \\ J_2^\top & J_3\end{bmatrix}\begin{bmatrix}\Omega_1 & C_2 & C_3 & D_2 \\ \mathsf{O}_{q\times n} & \mathsf{O}_{q\times n} & \mathsf{O}_{q\times n\rho} & I_q\end{bmatrix}\boldsymbol{\chi}(t).
\end{aligned} \tag{28}$$

To obtain a upper bound functional for $\dot{\mathsf{v}}(\boldsymbol{x}(t+\cdot)) - \mathsf{s}(\boldsymbol{z}(t), \boldsymbol{w}(t))$, assume $U \succ 0$, then applying (16) in Lemma 5 to the integral $-\int_{-r}^{0}\boldsymbol{x}^\top(t+\tau)U\boldsymbol{x}(t+\tau)\mathsf{d}\tau$ in (28) with the relation $M(\tau) = \boldsymbol{m}(\tau)\otimes I_n$ produces

$$\int_{-r}^{0}\boldsymbol{x}^\top(t+\tau)U\boldsymbol{x}(t+\tau)\mathsf{d}\tau \geq [*]\,(\mathsf{F}\otimes U)\left[\int_{-r}^{0}\boldsymbol{\varphi}(t+\tau)\mathsf{d}\tau\right]. \tag{29}$$

Merge (29) with the expression of $\dot{\mathsf{v}}(\boldsymbol{x}(t+\cdot)) - \mathsf{s}(\boldsymbol{z}(t), \boldsymbol{w}(t))$ in (28). Then the inequality

$$\dot{\mathsf{v}}(\boldsymbol{x}(t+\cdot)) - \mathsf{s}(\boldsymbol{z}(t), \boldsymbol{w}(t)) \leq \boldsymbol{\chi}^\top(t)\left(\boldsymbol{\Phi} + [*]\,J_1^{-1}\begin{bmatrix}\Omega_1 & C_2 & C_3 & D_2\end{bmatrix}\right)\boldsymbol{\chi}(t) \tag{30}$$

can be derived, where $\boldsymbol{\chi}(t)$ has been defined in (26). Specifically, the matrix $\boldsymbol{\Phi}$, which is denoted by

$$\begin{aligned}
\boldsymbol{\Phi} &:= \mathbf{Sy}\left(\begin{bmatrix} P & Q \\ \mathsf{O}_n & \mathsf{O}_{n\times n\rho} \\ Q^\top & R \\ \mathsf{O}_{q\times n} & \mathsf{O}_{q\times n\rho}\end{bmatrix}\begin{bmatrix}\Pi_1 & A_2 & A_3 & B_2 \\ M(0) & -M(-r) & -\widehat{\mathsf{M}} & \mathsf{O}_{n\rho\times q}\end{bmatrix}\right) \\
&+ \Big([S + rU]\oplus[-S]\oplus[-\mathsf{F}\otimes U]\oplus J_3\Big) + \mathbf{Sy}\left(\begin{bmatrix}\mathsf{O}_{(2n+n\rho)\times m} \\ J_2^\top\end{bmatrix}\begin{bmatrix}\Omega_1 & C_2 & C_3 & D_2\end{bmatrix}\right)
\end{aligned} \tag{31}$$

contains all the terms induced by $\dot{\mathsf{v}}(\boldsymbol{x}(t+\cdot)) - \mathsf{s}(\boldsymbol{z}(t), \boldsymbol{w}(t))$ in (28) excluding (25).

Based on the property of positive definite matrices, it is easy to see that if

$$\boldsymbol{\Phi} + [*]\,J_1^{-1}\begin{bmatrix}\Omega_1 & C_2 & C_3 & D_2\end{bmatrix} \prec 0, \;\; S \succ 0, \;\; U \succ 0 \tag{32}$$

is satisfied then the dissipative inequality in (14) : $\dot{\mathsf{v}}(\boldsymbol{x}(t+\cdot)) - \mathsf{s}(\boldsymbol{z}(t), \boldsymbol{w}(t)) \leq 0$ holds.

Moreover, knowing $J_1^{-1} \succ 0$ in (4) enables one to invoke Schur complement so that we have

$$\boldsymbol{\Psi} := \begin{bmatrix} \boldsymbol{\Phi} & \begin{bmatrix} \Omega_1 & C_2 & C_3 & D_2 \end{bmatrix}^\top \\ * & -J_1 \end{bmatrix} \prec 0, \; S \succ 0, \;\; U \succ 0, \tag{33}$$

holds if and only if (32) holds. As a result, it follows that (33) implies (14). By considering the properties of positive definite matrices, it is obvious that given $\boldsymbol{\Psi} \prec 0$ holds then $\exists \epsilon_3 < 0$: $\dot{\mathsf{v}}(\boldsymbol{x}(t+\cdot)) \leq \epsilon_3 \|\boldsymbol{x}(t)\|_2^2$. Since we have proved the existence of the lower bound for $\dot{\mathsf{v}}(\boldsymbol{x}(t+\cdot))$, we conclude that (33) implies both (13) and (14).

Clearly, $\boldsymbol{\Psi} \prec 0$ is nonconvex and hence it is not possible to perform congruence transformations due to the presence of the coupled products between $\Pi_1, A_2, A_3$ and $P, Q$. To obtain convex conditions, Lemma 2 is employed to circumvent the nonlinear terms in $\boldsymbol{\Psi}$ by introducing slack variables.

The following derivations are predicated on the methodologies in [41, 1, 4]. By observing the structures of the expressions (31) and (33), rewrite $\boldsymbol{\Psi} \prec 0$ into $\boldsymbol{\Psi} = \mathbf{Sy}\left[\mathbf{P}^\top \boldsymbol{\Pi}\right] + \widehat{\boldsymbol{\Phi}} \prec 0$, where

$$\mathbf{P} := \begin{bmatrix} P & \mathsf{O}_n & Q & \mathsf{O}_{n\times q} & \mathsf{O}_{n\times m} \end{bmatrix}, \; \boldsymbol{\Pi} := \begin{bmatrix} \Pi_1 & A_2 & A_3 & B_2 & \mathsf{O}_{n\times m} \end{bmatrix}, \tag{34}$$

and

$$\begin{aligned} \widehat{\boldsymbol{\Phi}} := \mathbf{Sy}&\left( \begin{bmatrix} Q \\ \mathsf{O}_{n\times n\rho} \\ R \\ \mathsf{O}_{(q+m)\times n\rho} \end{bmatrix} \begin{bmatrix} M(0) & -M(-r) & -\widehat{\mathsf{M}} & \mathsf{O}_{n\rho\times q+m} \end{bmatrix} \right) \\ &+ \Big( [S + rU] \oplus [-S] \oplus [-\mathsf{F} \otimes U] \oplus J_3 \oplus (-J_1) \Big) + \mathbf{Sy}\left( \begin{bmatrix} \mathsf{O}_{(2n+n\rho)\times m} \\ J_2^\top \\ I_m \end{bmatrix} \begin{bmatrix} \Omega_1 & C_2 & C_3 & D_2 & \mathsf{O}_m \end{bmatrix} \right) \end{aligned} \tag{35}$$

It is easy to observe that matrix $\widehat{\boldsymbol{\Phi}}$ contains the terms which have no product associations with respect to matrices $P$ and $Q$. Furthermore, $\boldsymbol{\Psi}$ can be reformulated into

$$\boldsymbol{\Psi} = \begin{bmatrix} * \end{bmatrix} \begin{bmatrix} \mathsf{O}_n & \mathbf{P} \\ * & \widehat{\boldsymbol{\Phi}} \end{bmatrix} \begin{bmatrix} \boldsymbol{\Pi} \\ I_{(2n+\rho n+q+m)} \end{bmatrix} \prec 0. \tag{36}$$

It is clear that (36) resembles one of the inequalities in (11) as part of the statements in Lemma 2. However, since there are two matrix inequalities in (11), a new matrix inequality must be constructed accordingly, having the expectation that no significant interference to the feasibility of the original inequality $\boldsymbol{\Psi} \prec 0$ will be introduced.

Consider

$$\Upsilon^\top \begin{bmatrix} \mathsf{O}_n & \mathbf{P} \\ * & \widehat{\boldsymbol{\Phi}} \end{bmatrix} \Upsilon \prec 0 \tag{37}$$

with $\Upsilon^\top := \begin{bmatrix} \mathsf{O}_{(q+m)\times(3n+\rho n)} & I_{q+m} \end{bmatrix}$, which can be further simplified into

$$\Upsilon^\top \begin{bmatrix} \mathsf{O}_n & \mathbf{P} \\ * & \widehat{\boldsymbol{\Phi}} \end{bmatrix} \Upsilon = \begin{bmatrix} J_3 + \mathbf{Sy}(D_2^\top J_2) & D_2^\top \\ * & -J_1 \end{bmatrix} \prec 0. \tag{38}$$

As a matter of fact, (38) is identical to the very matrix which is resulted from extracting the $2\times 2$ block matrix at the right bottom of $\boldsymbol{\Psi}$ or $\widehat{\boldsymbol{\Phi}}$. Consequently, one can conclude that (38) is automatically satisfied if condition

(36) holds. Since inequality (36) is equivalent to the inequality $\boldsymbol{\Psi} \prec 0$ in (33), the new constructed inequality (38) introduces no additional constraints to the original inequality $\boldsymbol{\Psi} \prec 0$.

To utilize Lemma 2, two matrices $\mathbf{U}, \mathbf{Y}$ need to be determined based on the inequalities in (11) which contain all bases of the null spaces of $\Upsilon$ and $\begin{bmatrix}\boldsymbol{\Pi}^\top & I_{(2n+\rho n+q+m)}\end{bmatrix}^\top$, respectively. In terms of $\Upsilon$, it is easy to conclude $\mathrm{rank}(\Upsilon) = q+m$, hence by the rank nullity theorem we have $\mathrm{rank}(\mathbf{U}) = 3n + \rho n$. Similarly, we can obtain $\mathrm{rank}(\mathbf{Y}) = n$. Without losing generality, we assume

$$\begin{gathered}\mathbf{Y} := \begin{bmatrix}-I_n & \boldsymbol{\Pi}\end{bmatrix}, \ \mathbf{U} := \begin{bmatrix}I_{3n+\rho n} & \mathsf{O}_{(3n+\rho n)\times(q+m)}\end{bmatrix}, \ \mathbf{Y}\mathbf{Y}_\perp = \mathbf{Y}\begin{bmatrix}\boldsymbol{\Pi} \\ I_{(2n+\rho n+q+m)}\end{bmatrix} = \mathsf{O}_{n\times(2n+\rho n+q+m)}, \\ \mathbf{U}_\perp = \Upsilon, \ \ \mathbf{U}\mathbf{U}_\perp = \mathsf{O}_{(3n+\rho n)\times(q+m)}\end{gathered} \tag{39}$$

which is compatible with the aforementioned rank values and stratifying the properties in Lemma 2.

Applying Lemma 2 to (36) and (38) yields that (36) and (38) are true if and only if

$$\exists \mathbf{W} \in \mathbb{R}^{(3n+\rho n)\times n} : \boldsymbol{\Theta} := \mathbf{Sy}\begin{bmatrix}\mathbf{U}^\top \mathbf{W}\mathbf{Y}\end{bmatrix} + \begin{bmatrix}\mathsf{O}_n & \mathbf{P} \\ * & \widehat{\boldsymbol{\Phi}}\end{bmatrix} \prec 0. \tag{40}$$

To ultimately produce convex synthesis conditions, consider

$$\mathbf{W} := \begin{bmatrix}W^\top & \eta_1 W^\top & \eta_2 W^\top & \widehat{W}^\top\end{bmatrix}^\top \tag{41}$$

where $W \in \mathbb{R}^{n\times n}_{[n]}$, $\widehat{W} := (\mathbf{col}_{i=1}^{\rho}\,\epsilon_i W) \in \mathbb{R}^{\rho n\times n}$, $\eta_1; \eta_2 \in \mathbb{R}$ and $\{\epsilon_i\}_{i=1}^{\rho} \in \mathbb{R}$, so that $\boldsymbol{\Theta} \prec 0$ can be convexified via congruence transformations. However, because of the structural constraints in (41), the corresponding (40) is no longer equivalent but only a sufficient condition implying (36) or $\boldsymbol{\Psi} \prec 0$.

Substituting both (41) and (39) into (40) produces the corresponding new $\boldsymbol{\Theta} \prec 0$. As a result, the original sufficient stability condition (33) is satisfied if

$$\boldsymbol{\Theta} \prec 0, \ \ S \succ 0, \ \ U \succ 0 \tag{42}$$

holds. Invoking congruence transformations on inequalities (42), and given the fact that $W$ is full rank, we have

$$\check{\mathbf{X}}^\top \boldsymbol{\Theta} \check{\mathbf{X}} \prec 0, \ \ X^\top S X \succ 0, \ \ X^\top U X \succ 0 \tag{43}$$

holds if and only if (42) holds, where $X^\top := W^{-1}$, $\check{\mathbf{X}} := X \oplus X \oplus X \oplus (I_\rho \otimes X) \oplus I_{q+m}$.

Moreover, letting

$$\check{P} := X^\top P X, \ \ \check{Q} := X^\top Q\,(I_\rho \otimes X)\,, \check{R} := (*)\,R\,(I_\rho \otimes X)\,, \ \ [\check{S} \ \ \check{U}] := X^\top [SX \ \ UX] \tag{44}$$

and using (8) to matrices $X$, $(I_\rho \otimes X)$ and $\widehat{\mathsf{M}}$ yields

$$\widehat{\mathsf{M}}\,(I_\rho \otimes X) = (I_\rho \otimes X)\,\widehat{\mathsf{M}}. \tag{45}$$

By taking both (45) and (44) into account, it yields

$$\begin{gathered}X^\top Q\widehat{\mathsf{M}}\,(I_\rho \otimes X) = X^\top Q\,(I_\rho \otimes X)\,\widehat{\mathsf{M}} = \check{Q}\widehat{\mathsf{M}}, \\ (I_\rho \otimes X)^\top R\widehat{\mathsf{M}}\,(I_\rho \otimes X) = (I_\rho \otimes X)^\top R\,(I_\rho \otimes X)\,\widehat{\mathsf{M}} = \check{R}\widehat{\mathsf{M}}, \\ (I_\rho \otimes X)^\top [\mathsf{F} \otimes S \ \ \ \mathsf{F} \otimes U]\,(I_{2\rho} \otimes X) = \begin{bmatrix}\mathsf{F} \otimes \check{S} & \mathsf{F} \otimes \check{U}\end{bmatrix}.\end{gathered} \tag{46}$$

Furthermore, let $V = KX$ and consider all the relations in (44) and (3). Then $\check{\mathbf{X}}^\top \boldsymbol{\Theta} \check{\mathbf{X}} \prec 0$ in (43) becomes (22). This proves the equivalence between (22) and $\boldsymbol{\Theta} \prec 0$ as $\check{\mathbf{X}}^\top \boldsymbol{\Theta} \check{\mathbf{X}} \prec 0$ is equivalent to $\boldsymbol{\Theta} \prec 0$. Due to the fact that the expression $-X - X^\top$ is the only element in the first block diagonal $\check{\mathbf{\Theta}}$, thus $X$ cannot be singular if $\check{\mathbf{\Theta}} \prec 0$ holds. Consequently, given (22) with $\check{S} \succ 0, \check{U} \succ 0$ implies both (13) and (14) which proves that the closed loop system (3) is dissipative with respect to (4). This finishes the proof of *Part 1*. From onwards, we start to prove *Part 2*.

*Part 2.*

Given $\check{S} \succ 0$ in (21) and considering $\check{S} = X^\top S X$ and $X \in \mathbb{R}^{n\times n}_{[n]}$ in (44) implies $S \succ 0$. Subsequently, applying (16) with $M(\tau) = \boldsymbol{m}(\tau) \otimes I_n$ to the integral $\int_{-r}^{0} \boldsymbol{\varphi}^\top(t+\tau) S \boldsymbol{\varphi}(t+\tau) \mathrm{d}\tau$ yields

$$\int_{-r}^{0} \boldsymbol{x}^\top(t+\tau) S \boldsymbol{x}(t+\tau) \mathrm{d}\tau \geq \left[\ast\right] (\mathsf{F} \otimes S) \int_{-r}^{0} \boldsymbol{\varphi}(t+\tau)\mathrm{d}\tau. \tag{47}$$

Now we merge (47) with (20), so that a new functional

$$\mathsf{v}_1(\boldsymbol{x}(t+\cdot)) := \left[\ast\right] \begin{bmatrix} P & Q \\ \ast & R + \mathsf{F}\otimes S \end{bmatrix} \begin{bmatrix} \boldsymbol{x}(t) \\ \int_{-r}^{0} \boldsymbol{\varphi}(t+\tau)\mathrm{d}\tau \end{bmatrix} + \int_{-r}^{0} (\tau + r) \boldsymbol{x}^\top(t+\tau) U \boldsymbol{x}(t+\tau) \mathrm{d}\tau \tag{48}$$

can be obtained to provide a lower bound for $\mathsf{v}(\boldsymbol{x}(t+\cdot))$ such that $\mathsf{v}(\boldsymbol{x}(t+\cdot)) \geq \mathsf{v}_1(\boldsymbol{x}(t+\cdot))$.

According to the property of positive definite matrices and all the previous derived conclusions, it is obvious to conclude that if

$$\begin{bmatrix} P & Q \\ \ast & R + \mathsf{F}\otimes S \end{bmatrix} \succ 0, \;\; S \succ 0, \;\; U \succ 0 \tag{49}$$

is satisfied, then $\exists \epsilon_1 > 0 : \mathsf{v}_1(\boldsymbol{x}(t+\cdot)) \geq \epsilon_1 \|\boldsymbol{x}(t)\|_2$ which implies $\exists \epsilon_1 > 0 : \mathsf{v}(\boldsymbol{x}(t+\cdot)) \geq \epsilon_1 \|\boldsymbol{x}(t)\|_2$ since $\mathsf{v}(\boldsymbol{x}(t+\cdot)) \geq \mathsf{v}_1(\boldsymbol{x}(t+\cdot))$.

Invoking congruence transformations to some of the inequalities in (49) and given $X$ is nonsingular, we can conclude that (49) holds if and only if

$$(\ast) \begin{bmatrix} P & Q \\ \ast & R + \mathsf{F}\otimes S \end{bmatrix} (X \oplus (I_\rho \otimes X)) \succ 0, \;\; X^\top U X \succ 0, \;\; \check{S} \succ 0, \tag{50}$$

holds. By using the relations in (44) and (3) further, (50) becomes (21), thus (21) is proved to be equivalent to (49) as (50) is equivalent to (49). Furthermore, it is obvious that $\forall X \in \mathbb{S}^n, \exists \lambda > 0 : \forall \mathbf{x} \in \mathbb{R}^n \backslash \{\mathbf{0}\}, \mathbf{x}^\top (\lambda I_n - X)\mathbf{x} > 0$. Consequently, it follows that $\exists \lambda_1, \lambda_2 > 0$ such that

$$\begin{aligned} \mathsf{v}(\boldsymbol{x}(t+\cdot)) &\leq [\ast]\, \lambda_1 \begin{bmatrix} \boldsymbol{x}(t) \\ \int_{-r}^{0} \boldsymbol{\varphi}(t+\tau)\mathrm{d}\tau \end{bmatrix} \mathrm{d}\tau + \int_{-r}^{0} r\lambda_1 \sup_{\tau\in\mathcal{O}} \|\boldsymbol{x}(t+\tau)\|_2^2 \mathrm{d}\tau \\ &\leq [\ast]\lambda_1 \boldsymbol{x}(t) + r^2 \lambda_1 \sup_{\tau\in\mathcal{O}} \|\boldsymbol{x}(t+\tau)\|_2^2 + \lambda_1 \int_{-r}^{0} \boldsymbol{\varphi}^\top(t+\tau)(\mathsf{F}\otimes I_n)(\mathsf{F}^{-1}\otimes I_n)\boldsymbol{\varphi}(t+\tau)\mathrm{d}\tau \\ &\leq [\ast]\lambda_1 \boldsymbol{x}(t) + \int_{-r}^{0} [\ast]\lambda_2 \boldsymbol{x}(t+\tau)\mathrm{d}\tau + r^2\lambda_1 \sup_{\tau\in\mathcal{O}} \|\boldsymbol{x}(t+\tau)\|_2^2 \leq \left(\lambda_1 + r\lambda_2 + r^2\lambda_1\right) \sup_{\tau\in\mathcal{O}} \|\boldsymbol{x}(t+\tau)\|_2^2, \end{aligned} \tag{51}$$

which is derived via the application of (16).

Thus we have proved that $\exists \epsilon_2 > 0 : \mathsf{v}(\boldsymbol{x}(t+\cdot)) \leq \epsilon_2 \sup_{\tau\in\mathcal{O}} \|\boldsymbol{x}(t+\tau)\|_2^2$. Consequently, it follows that (21) implies (12), as (21) has been proved to be equivalent to (49). This finishes the proof of *Part 2*.

According to the conclusion from both *Part 1* and *Part 2*, the conditions (21) and (22) imply the original stability criteria (12) and (13) satisfying the dissipative inequality in (14). This completes the proof. ■

**Remark 3.** For conducting a standard stability analysis incorporating dissipativity for the open loop system (1), there is no need to apply the Projection Lemma as (33) and (49) are already convex without considering the controller gains $B_1$ and $D_1$. As a result, we have not set up a theorem specifically for stability analyses. Nevertheless, one can easily apply (49) and (33) as stability conditions for the open loop system (1). Furthermore, for conducting a simple analysis without considering supply functions and the output signal $\boldsymbol{y}(t)$, one can apply

$$\boldsymbol{\Phi} := \mathbf{Sy}\left(\begin{bmatrix} P & Q \\ \mathsf{O}_n & \mathsf{O}_{n\times n\rho} \\ Q^\top & R \end{bmatrix}\begin{bmatrix} \Pi_1 & A_2 & A_3 \\ M(0) & -M(-r) & -\widehat{\mathsf{M}} \end{bmatrix}\right) + \Big([S+rU]\oplus[-S]\oplus[-\mathsf{F}\otimes U]\Big) \succ 0, \tag{52}$$

with $S \succ 0$, $U \succ 0$ and (49) to attain this task.

**Remark 4.** Even without considering dissipativity constraints, it is still possible to introduce slack variables as in (40) to conduct a stability analysis. However, in such situation, the Projection Lemma is not an option since it is not possible to construct a matrix inequality as illustrated in (38). Instead, a particular version of the Projection Lemma which only demands one inequality as in (11), called the Finsler Lemma [6], can be applied to handle this problem. To be specific, a standard negative stability condition without considering dissipativity is expressed as

$$\mathbf{Sy}\left(\mathbf{WY}\right) + \begin{bmatrix} \mathsf{O}_n & \mathbf{P} \\ * & \widehat{\boldsymbol{\Phi}} \end{bmatrix} \prec 0, \tag{53}$$

where $\mathbf{W} \in \mathbb{R}^{(3n+\rho n)\times n}$ with the matrices $\mathbf{P}$ and $\widehat{\boldsymbol{\Phi}}$ deleting all the terms associated with performance characterizations.

**Remark 5.** It is important to stress that one can simply apply the Finsler lemma to (36) so that a similar equality with more extra variables than $\mathbf{W}$ in (40) can be obtained, as it has been demonstrated in (53). However, by comparing (53) and (40), it is obvious that the numbers of the slack variable $\mathbf{W}$ are identical. This indicates the fact that the new constructed inequality (38) functions as simplifying the numbers of decision variable and at the same time introducing no extra conservatism. Consequently, this can be regarded as another reason to utilize Projection Lemma in this paper apart from achieving linearizations only.

### *3.1. Controller synthesis of uncertain linear delay systems with dissipativity constraints*

The aforementioned design scheme can be easily extended to handle open loop linear delay systems with general uncertainties. In the following derivations, we assume that all the state space matrices in (1) exhibit linear fractional uncertainty structures :

$$\begin{aligned}&\begin{bmatrix} \grave{A}_1 & \grave{B}_1 & \grave{A}_2 & \grave{A}_3 & \grave{B}_2 & \grave{C}_1 & \grave{D}_1 & \grave{C}_2 & \grave{C}_3 & \grave{D}_2 \end{bmatrix} \\ &\qquad = \begin{bmatrix} A_1 & B_1 & A_2 & A_3 & B_2 & C_1 & D_1 & C_2 & C_3 & D_2 \end{bmatrix} + \mathop{\mathbf{row}}_{i=1}^{10}\left[G_i(I_? - \Delta_i F_i)^{-1}\Delta H_i\right]\end{aligned} \tag{54}$$

with the full block constraints in [31, 4]

$$\Delta_i \in \left\{ \widehat{\Delta}_i \;\middle|\; \begin{bmatrix} I \\ \widehat{\Delta}_i \end{bmatrix}^\top \begin{bmatrix} \Xi_i^{-1} & \Lambda_i \\ * & \Gamma_i \end{bmatrix}\begin{bmatrix} I \\ \widehat{\Delta}_i \end{bmatrix} \succeq 0 \right\}, \;\; \forall i = 1\cdots 10, \;\; \Xi_i^{-1} \succ 0, \;\; \Gamma_i \preceq 0, \tag{55}$$

where all matrices are supposed to have compatible dimensions. Moreover, the question marks ? in (54) are determined by the specific dimensions of $\Delta_i, i = 1\cdots 10$ and $F_i, i = 1\cdots 10$.

The constraints of uncertainty (55) can be rewritten into $\widehat{\Delta}_i^\top\Gamma_i\widehat{\Delta}_i + \mathbf{Sy}(\Lambda_i\widehat{\Delta}_i) + \Xi_i^{-1} \succeq 0, \forall i = 1\cdots 10$, which is equivalent to a single inequality $\bigoplus_{i=1}^{10}\left(\widehat{\Delta}_i^\top\Gamma_i\widehat{\Delta}_i + \mathbf{Sy}(\Lambda_i\widehat{\Delta}_i) + \Xi_i^{-1}\right) \succeq 0$. Moreover, using the property of diagonal matrices $(X+Y)\oplus(X+Y) = (X\oplus X) + (Y\oplus Y)$ and $XY\oplus XY = (X\oplus X)(Y\oplus Y)$ yields

$$\bigoplus_{i=1}^{10}\Delta_i \in \mathcal{U} := \left\{ \bigoplus_{i=1}^{10}\widehat{\Delta}_i \;\middle|\; [*]\begin{bmatrix} \bigoplus_{i=1}^{10}\Xi_i^{-1} & \bigoplus_{i=1}^{10}\Lambda_i \\ * & \bigoplus_{i=1}^{10}\Gamma_i \end{bmatrix}\begin{bmatrix} I \\ \bigoplus_{i=1}^{10}\widehat{\Delta}_i \end{bmatrix} \succeq 0 \right\}, \;\; \bigoplus_{i=1}^{10}\Xi_i^{-1} \succ 0, \;\; \bigoplus_{i=1}^{10}\Gamma_i \preceq 0. \tag{56}$$

Having demonstrated the equivalence relation between (56) and (55), (56) will be applied in deriving the robust synthesis conditions due to its compact structure.

**Remark 6.** The expression in (54) with the constraints in (56) provides nearly a complete characterization in terms of the framework of norm bounded uncertainties. In addition, understand that the robust terms $\grave{A}_3 = A_3 + G_4(I_? - \Delta_4 F_4)^{-1}\Delta H_4$ and $\grave{C}_3 = C_3 + G_9(I_? - \Delta_9 F_9)^{-1}\Delta H_9$ lead to the distributed terms $\widetilde{A}_3(\tau) = A_3 M(\tau) + G_4(I_? - \Delta_4 F_4)^{-1}\Delta H_4 M(\tau)$ and $\widetilde{C}_3(\tau) = C_3 M(\tau) + G_9(I_? - \Delta_9 F_9)^{-1}\Delta H_9 M(\tau)$, respectively. This further demonstrates the fact that the uncertainties associated with the distributed terms are sufficiently general, as all the coefficients of the functions in $\widetilde{A}_3(\tau), \widetilde{C}_3(\tau)$ are subject to the variations of $G_4(I_? - \Delta_4 F_4)^{-1}\Delta H_4$ and $G_9(I_? - \Delta_9 F_9)^{-1}\Delta H_9$, respectively.

In order to handle the uncertainties structure in (54) and (56), the following Lemma is derived.

**Lemma 6.** *For arbitrary* $n; m; p; q \in \mathbb{N}$, $\Theta_1 \in \mathbb{S}^p_{\succ 0}$, $\Theta_3 \in \mathbb{S}^m_{\preceq 0}$, $\Theta_2 \in \mathbb{R}^{p\times m}$, $\Phi \in \mathbb{S}^n$, $G \in \mathbb{R}^{n\times m}$, $H \in \mathbb{R}^{p\times n}$, $F \in \mathbb{R}^{p\times m}$ *if*

$$\exists \alpha > 0 : \begin{bmatrix} I_m & -I_m - \alpha F^\top \Theta_2 & \alpha F^\top \\ * & I_m - \alpha\Theta_3 & \mathsf{O}_{m\times p} \\ * & * & \alpha\Theta_1 \end{bmatrix} \succ 0, \tag{57}$$

*then we have the inequality*

$$\Phi + \mathbf{Sy}\left[G(I_m - \Delta F)^{-1}\Delta H\right] \prec 0,\ \forall \Delta \in \mathcal{F} \subseteq \mathcal{D} := \left\{ \widehat{\Delta} \in \mathbb{R}^{m\times p} \,\middle|\, [*]\begin{bmatrix} \Theta_1^{-1} & \Theta_2 \\ * & \Theta_3 \end{bmatrix}\begin{bmatrix} I_p \\ \widehat{\Delta} \end{bmatrix} \succeq 0 \right\} \tag{58}$$

*holds provided that*

$$\exists \kappa > 0 : \begin{bmatrix} \Phi & G + \kappa H^\top \Theta_2 & \kappa H^\top \\ * & \kappa F^\top \Theta_2 + \kappa \Theta_2^\top F + \Theta_3 & \kappa F^\top \\ * & * & -\kappa\Theta_1 \end{bmatrix} \prec 0. \tag{59}$$

*For the situation when* $\Theta_1^{-1} = \mathsf{O}_p$, (59) *and* (57) *become*

$$\exists \kappa > 0 : \begin{bmatrix} \Phi & G + \kappa H^\top \Theta_2 \\ * & \kappa F^\top \Theta_2 + \kappa \Theta_2^\top F + \kappa\Theta_3 \end{bmatrix} \prec 0, \tag{60}$$

$$\exists \alpha > 0 : \begin{bmatrix} I_m & -I_m - \alpha F^\top \Theta_2 \\ * & I_m - \alpha\Theta_3 \end{bmatrix} \succ 0, \tag{61}$$

*respectively.*

*Proof.* see Appendix B for details. It is important to mention that (58) and (59) become equivalent if $\mathcal{F} = \mathcal{D}$. ■

**Remark 7.** It is vital to stress that the result in Lemma 6 is not identical to the conclusions in [31], given the fact that $\mathcal{F}$ does not need to be compact. The assumption $\Theta_1 \succ 0$ in this lemma is motivated by the demand to produce convex conditions considering the structure of (59) with a fixed $\kappa$.

The aforementioned lemma is able to cover a wide range of uncertainty configurations such as the common norm bounded uncertainties: $\Theta_1^{-1} = R \succ 0$, $\Theta_2 = \mathsf{O}_{p\times m}$, $\Theta_3 = -I_m$, $F = \mathsf{O}_{p\times m}$. Furthermore, when $F \neq \mathsf{O}_{p\times m}$, $\Theta_1^{-1} = I_p$, $\Theta_2 = \mathsf{O}_{p\times m}$, $\Theta_3 = -I_m$, it becomes the linear fractional uncertainties demonstrated in [40], provided that the corresponding well-posedness condition (57) is satisfied.

Now we combine the synthesis results in Theorem 1 with (54) and (56). By using the conclusion in Lemma 4, it results in the following theorem which provides sufficient conditions for the existence of a state feedback controller ensuring both dissipativity and robustness.

**Theorem 2.** *Given all conditions to be held in Theorem 1, the closed loop system* (3) *incorporating* (54) *is robustly globally asymptotically stable* ($\boldsymbol{w}(t) \equiv \mathbf{0}_q$) *at the origin within the uncertainty set* $\mathcal{U}$ *in* (56) *and dissipative with the supply function* (4), *if there exist* $\varkappa_1, \varkappa_2 > 0$ *such that the following conditions are satisfied,*

$$\begin{bmatrix} I & -I - \varkappa_1 \mathbf{F}^\top \mathbf{J}_2 & \varkappa_1 \mathbf{F}^\top \\ * & I - \varkappa_1 \mathbf{J}_3 & \mathsf{O} \\ * & * & \varkappa_1 \mathbf{J}_1 \end{bmatrix} \succ 0, \quad \begin{bmatrix} \check{\mathbf{\Theta}} & \mathbf{G} + \varkappa_2 \mathbf{H}^\top \mathbf{J}_2 & \varkappa_2 \mathbf{H}^\top \\ * & \varkappa_2 \mathbf{F}^\top \mathbf{J}_2 + \varkappa_2 \mathbf{J}_2^\top \mathbf{F} + \varkappa_2 \mathbf{J}_3 & \varkappa_2 \mathbf{F}^\top \\ * & * & -\varkappa_2 \mathbf{J}_1 \end{bmatrix} \prec 0, \tag{62}$$

*where* $\check{\mathbf{\Theta}}$ *is defined in Theorem 1 and*

$$\mathbf{G} := \begin{bmatrix} I_n & \mathsf{O}_{n\times m} \\ \eta_1 I_n & \mathsf{O}_{n\times m} \\ \eta_2 I_n & \mathsf{O}_{n\times m} \\ \mathbf{col}_{i=1}^{\rho}\, \epsilon_i I_n & \mathsf{O}_{\rho n\times m} \\ \mathsf{O}_{q\times n} & J_2^\top \\ \mathsf{O}_{m\times n} & I_m \end{bmatrix} \left( \left[ \mathop{\mathbf{row}}_{i=1}^{5} G_i \right] \oplus \left[ \mathop{\mathbf{row}}_{i=6}^{10} G_i \right] \right), \tag{63}$$

$$\mathbf{col}\left[\mathbf{F}, \mathbf{J}_1, \mathbf{J}_2, \mathbf{J}_3\right] := \mathbf{col}\left[ \bigoplus_{i=1}^{10} F_i, \bigoplus_{i=1}^{10} \Xi_i, \bigoplus_{i=1}^{10} \Lambda_i, \bigoplus_{i=1}^{10} \Gamma_i \right] \tag{64}$$

$$\mathbf{H} := \begin{pmatrix} \begin{bmatrix} H_1 X \\ H_2 V \end{bmatrix} \oplus H_3 X \oplus [H_4 (I_\rho \otimes X)] \oplus H_5 \\ \begin{bmatrix} H_6 X \\ H_7 V \end{bmatrix} \oplus H_8 X \oplus [H_9 (I_\rho \otimes X)] \oplus H_{10} \end{pmatrix} \begin{bmatrix} \mathsf{O}_{n\times(2n+n\rho+q)} \\ I_{2n+n\rho+q} \\ \mathsf{O}_{m\times(2n+n\rho+q)} \end{bmatrix}^\top \tag{65}$$

*Proof.* Substituting (54) into (3) and decomposing the corresponing (22) into

$$\begin{aligned} &\check{\mathbf{\Theta}} + \begin{bmatrix} I_n & \mathsf{O}_{n\times m} \\ \eta_1 I_n & \mathsf{O}_{n\times m} \\ \eta_2 I_n & \mathsf{O}_{n\times m} \\ \mathop{\mathbf{col}}_{i=1}^{\rho} \epsilon_i I_n & \mathsf{O}_{\rho n\times m} \\ \mathsf{O}_{q\times n} & J_2^\top \\ \mathsf{O}_{m\times n} & I_m \end{bmatrix} \begin{bmatrix} \mathsf{O}_n & (\grave{A}_1 - A_1)X + (\grave{B}_1 - B_1)V & \mathop{\mathbf{row}}_{i=2}^{3}(\grave{A}_i - A_i)X & (\grave{B}_2 - B_2) & \mathsf{O}_{n\times m} \\ \mathsf{O}_{m\times n} & (\grave{C}_1 - C_1)X + (\grave{D}_1 - D_1)V & \mathop{\mathbf{row}}_{i=2}^{3}(\grave{C}_i - C_i)X & (\grave{D}_2 - D_2) & \mathsf{O}_m \end{bmatrix} \\ &= \check{\mathbf{\Theta}} + \mathbf{Sy}\left[ \mathbf{G} \left( \mathbf{I} - \mathbf{\Delta F} \right)^{-1} \mathbf{\Delta H} \right] \prec 0, \end{aligned} \tag{66}$$

where $\mathbf{\Delta} := \bigoplus_{i=1}^{10} \Delta_i$ is compatible with the constraints in (56) and the matrices $\mathbf{G}, \mathbf{H}, \mathbf{F}$ have been defined in (63) and (65), respectively. The resulting decomposition can be easily derived based on the structure seen in (22) and (23), which exhibits a symmetric distribution with respect to the matrix terms associated with uncertainties.

Applying the result in Lemma 4 to (66) with (56) yields the condition (62). This completes the proof. ■

**Remark 8.** It is worthy to mention that all the uncertainties in (54) can be 'pulled out' into interconnection form as it has been demonstrated in [29, 31]. However, for the sake of producing a single convex condition not requiring the application of the dualization Lemma [29, 4], the uncertainties have been chosen in the form of (54).

**Remark 9.** Motivated by the reasons explained in Remark 3, we have not established a theorem concerning the robust stability analyses of an open loop system such as (1). However, the conditions for stability analyses can be easily derived by considering the inequality

$$\mathbf{\Psi} + \begin{bmatrix} P & \mathsf{O}_{n\times m} \\ \mathsf{O} & \mathsf{O}_{n\times m} \\ Q^\top & \mathsf{O}_{\rho n\times m} \\ \mathsf{O}_{q\times n} & J_2^\top \\ \mathsf{O}_{m\times n} & I_m \end{bmatrix} \begin{bmatrix} \mathop{\mathbf{row}}_{i=1}^{3}(\grave{A}_i - A_i) & \grave{B}_2 - B_2 & \mathsf{O}_{n\times m} \\ \mathop{\mathbf{row}}_{i=1}^{3}(\grave{C}_i - C_i) & \grave{D}_2 - D_2 & \mathsf{O}_m \end{bmatrix} = \mathbf{\Psi} + \mathbf{Sy}\left[ \mathbf{G} \left( \mathbf{I} - \mathbf{\Delta F} \right)^{-1} \mathbf{\Delta H} \right] \prec 0, \tag{67}$$

where $\mathbf{\Psi}$ has been defined in (33) and

$$\mathbf{G} := \begin{bmatrix} P & \mathsf{O}_{n\times m} \\ \mathsf{O}_n & \mathsf{O}_{n\times m} \\ Q^\top & \mathsf{O}_{\rho n\times m} \\ \mathsf{O}_{q\times n} & J_2^\top \\ \mathsf{O}_{m\times n} & I_m \end{bmatrix} \left( \begin{bmatrix} G_1 & \mathbf{row}_{i=3}^5 G_i \end{bmatrix} \oplus \begin{bmatrix} G_6 & \mathbf{row}_{i=8}^{10} G_i \end{bmatrix} \right),$$

$$\begin{bmatrix} \mathbf{F} \\ \mathbf{J}_1 \\ \mathbf{J}_2 \\ \mathbf{J}_3 \end{bmatrix} := \begin{bmatrix} F_1 \oplus \bigoplus_{i=3}^{6} F_i \oplus \bigoplus_{i=8}^{10} F_i \\ \Xi_1 \oplus \bigoplus_{i=3}^{6} \Xi_i \oplus \bigoplus_{i=8}^{10} \Xi_i \\ \Lambda_1 \oplus \bigoplus_{i=3}^{6} \Lambda_i \oplus \bigoplus_{i=8}^{10} \Lambda_i \\ \Gamma_1 \oplus \bigoplus_{i=3}^{6} \Gamma_i \oplus \bigoplus_{i=8}^{10} \Gamma_i \end{bmatrix}, \quad \mathbf{H} := \begin{bmatrix} H_1 \oplus \bigoplus_{i=3}^5 H_i, \\ H_6 \oplus \bigoplus_{i=8}^{10} H_i \end{bmatrix} \begin{bmatrix} I_{2n+n\rho+q} & \mathsf{O}_{(2n+n\rho+q)\times m} \end{bmatrix}. \tag{68}$$

in accordance to the expressions in (54) and (56). As a result, applying Lemma 6 to (67) with (56) will yield the corresponding robust stability condition which will be similar to (62). Finally, it should be pointed out that (49) remains unchanged as part of the stability conditions even with the consideration of the uncertainties in (54).

## 4. Numerical Simulations

The following numerical examples are tested in Matlab with Yalmip [20] and [37] as the optimization interface and the solver for semidefinite programming, respectively. Furthermore, all the analytic properties of the delay systems considered in this section are examined by the methodology proposed in [2] with the codes in [3].

### *4.1. Stability analysis of distribtuted delay systems*

Consider a distributed delay system

$$\dot{x}(t) = 0.395x(t) - 5\int_{-r}^{0} \cos(12\tau)x(t+\tau)\mathrm{d}\tau. \tag{69}$$

Since the corresponding delay free system is unstable and distributed term contains a trigonometric function, the methodologies in [27] and [34] are not able to produce any feasible result.

In order to apply the methodology in this paper, $M(\tau)$ is chosen to be

$$M(\tau) = \underbrace{\begin{bmatrix} 1 \\ \sin(12\tau) \\ \cos(12\tau) \end{bmatrix}}_{m(\tau)} \otimes 1 \quad \text{with} \quad \widehat{\mathsf{M}} = \underbrace{\begin{bmatrix} 0 & 0 & 0 \\ 0 & 0 & 12 \\ 0 & -12 & 0 \end{bmatrix}}_{\mathsf{M}} \otimes 1 \tag{70}$$

with $A_3 = \begin{bmatrix} 0 & 0 & -5 \end{bmatrix}$. As a result, we have $\widetilde{A}_3(\tau) = A_3 M(\tau)$ with $\rho = 3, n = 1$ which satisfies Assumption 1.

Furthermore, applying the spectrum methods in [2] with 20 as the discretization index yields Figure 1 as a stability diagram, where it plots the values of sign $[\Re(\lambda)]$ such that $\lambda$ denotes the rightmost characteristics roots of the system (69). Specifically, the only stable intervals (this is verified by testing sufficient large $r$) of the delay system are $[0.104, 0.1578]$, $[0.6276, 0.6814]$, $[1.1512, 1.205]$, $[1.6748, 1.7286]$ and $[2.1984, 2.2522]$.

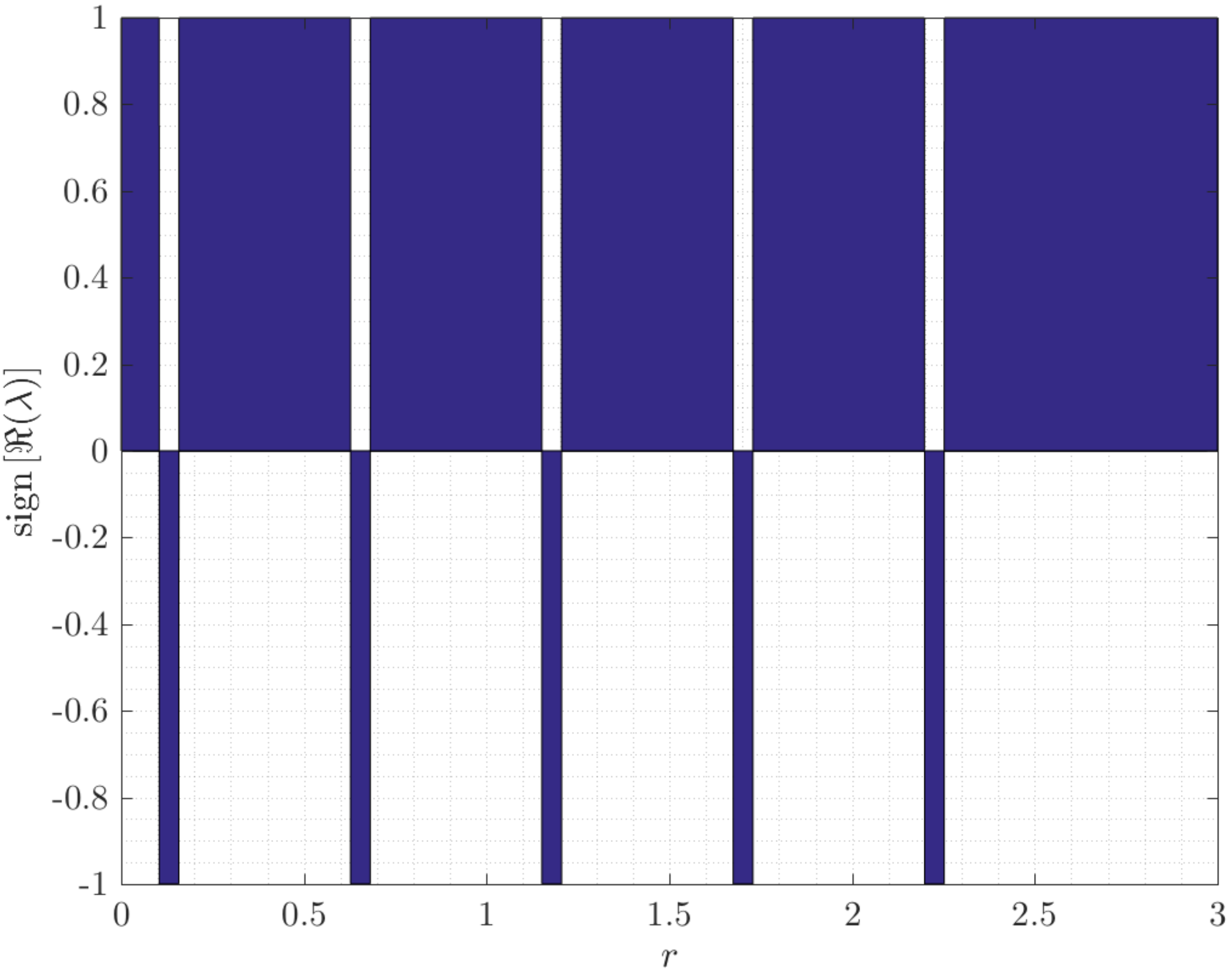

Figure 1: Diagram showing stability regions of (69)

By utilizing the methodology proposed in [35] to (69), it produces the results showing in Table 2. It can be observed that it requires $\rho = 3$ (which is equivalent to the $N$ in [35]) with 12 variables to detect all the boundaries of the first stability intervals. As $r$ increases, a larger degree of the approximation polynomials $\rho$ is demanded in order to produce feasible results. It follows that $\rho = 22$ with 302 variables id required to detect the upper stability boundary 2.2522.

In contrast, applying the conditions (49) and (52) with (70) to (69), we are able to obtain all the stability interval with only 12 decision variables. It is important to stress that each time the matrix $\mathsf{F} = \left[\int_{-r}^{0} \boldsymbol{m}(\tau)\boldsymbol{m}^\top(\tau)\mathsf{d}\tau\right]^{-1}$ in Theorem 1 has to be calculated in accordance to the value of $r$.

**Remark 10.** The instability intervals can be accurately detected by the approach in [35] and the proposed methodology, as the eigenvalues of some positive definite decision variables immediately turn into negatives when $r$ exceeds the stability boundaries. This illustrates the fact that the results from both [35] and this paper are consistent with the reliable calculations in [2], which is indeed not common in comparison to existing time domain approaches. However, a clear contribution of this paper is rooted in that less variables might be required for the distributed kernels exhibiting patterns of intensive oscillations, which has been demonstrated by the aforementioned example. Finally, it is worth noting that the aforementioned stability results can be accurately reproduced, however, with more variables, by the inequalities concerning slack variables in (53). This further ensures that the synthesis approach in this paper by means of slack variables is able to produce reliable results.

| Methodology | first interval | second interval | third interval | forth interval | fifth interval | NDVs |
|---|---|---|---|---|---|---|
| [34] | – | – | – | – | – | – |
| [26] | – | – | – | – | – | – |
| $\rho = 2$, [35] | $[0.104, 0.1578]$ | – | – | – | – | 12 |
| $\rho = 8$, [35] | $[0.104, 0.1578]$ | $[0.6276, 0.6814]$ | – | – | – | 57 |
| $\rho = 13$, [35] | $[0.104, 0.1578]$ | $[0.6276, 0.6814]$ | $[1.1512, 1.205]$ | – | – | 122 |
| $\rho = 17$, [35] | $[0.104, 0.1578]$ | $[0.6276, 0.6814]$ | $[1.1512, 1.205]$ | $[1.6748, 1.7286]$ | – | 192 |
| $\rho = 22$, [35] | $[0.104, 0.1578]$ | $[0.6276, 0.6814]$ | $[1.1512, 1.205]$ | $[1.6748, 1.7286]$ | $[2.1984, 2.2522]$ | 302 |
| $\rho = 2$, our approach | $[0.104, 0.1578]$ | $[0.6276, 0.6814]$ | $[1.1512, 1.205]$ | $[1.6748, 1.7286]$ | $[2.1984, 2.2522]$ | 12 |

Figure 2: Feasible Stability Testing Intervals (NDV stands for the number of decision variables).

### 4.2. Controller Synthesis

#### 4.2.1. Synthesis considering dissipativity

Consider a system of the form (1) with a constant delay $r = 1$ and state space matrices

$$A_1 = \begin{bmatrix} 0 & 0 \\ 0 & 0.1 \end{bmatrix},\ A_2 = \begin{bmatrix} -1 & -1 \\ 0 & 0.9 \end{bmatrix},\ B_1 = \begin{bmatrix} 0 \\ 1 \end{bmatrix},\ B_2 = \begin{bmatrix} 0.1 & -0.11 \\ 0.21 & 0.1 \end{bmatrix},\ D_1 = \begin{bmatrix} 0.2 \\ 0.3 \end{bmatrix},$$

$$\widetilde{A}_3(\tau) = \begin{bmatrix} -0.4-0.1e^{\tau}\sin(20\tau)+0.3e^{\tau}\cos(20\tau) & 1+0.2e^{\tau}\sin(20\tau)+0.2e^{\tau}\cos(20\tau) \\ -1+0.01e^{\tau}\sin(20\tau)-0.2e^{\tau}\cos(20\tau) & 0.4+0.3e^{\tau}\sin(20\tau)+0.4e^{\tau}\cos(20\tau) \end{bmatrix},\ D_2 = \begin{bmatrix} 0.1 & 0.2 \\ 0.12 & 0.1 \end{bmatrix} \tag{71}$$

$$C_1 = \begin{bmatrix} -0.1 & 0.2 \\ 0 & 0.1 \end{bmatrix},\ C_2 = \begin{bmatrix} -0.1 & 0 \\ 0 & 0.2 \end{bmatrix},\ \widetilde{C}_3(\tau) = \begin{bmatrix} 0.2e^{\tau}\sin(20\tau) & 0.1+0.1e^{\tau}\cos(20\tau) \\ 0.1e^{\tau}\sin(20\tau)-0.1e^{\tau}\cos(20\tau) & -0.2+0.3e^{\tau}\sin(20\tau) \end{bmatrix},$$

Since $(A_1, B_1)$ is not controllable, the stabilization method in [26] cannot be applied here regardless of the fact that $\widetilde{A}_3(\tau)$ and $\widetilde{C}_3(\tau)$ can be approximated via rational functions. Moreover, based on the structures of $\widetilde{A}_3(\tau)$ and $\widetilde{C}_3(\tau)$, the corresponding delay system does not have either forwarding or backstepping structures without having transformations. Hence, the constructive approaches in [17] may not be applicable here.

By using the spectrum method in [2] to (1) with the parameters (71), it occurs that the system is unstable for $0 \le r \le 10$. In terms of specifying a controller objective, let $J_1 = -J_3 = \gamma I_2,\ J_2 = \mathsf{O}_2$ for the supply rate function in (4) to calculate the minimum value of $\mathcal{L}_2$ attenuation factor $\gamma$ given that $\boldsymbol{w}(\cdot) \in \mathcal{L}_2(\mathbb{T}; \mathbb{R}^q_{\|2})$.

By observing the elements inside of $\widetilde{A}_3(\tau), \widetilde{C}_3(\tau)$, we choose $M(\tau)$ to be

$$M(\tau) = \underbrace{\begin{bmatrix} 1 \\ 10e^{\tau}\sin(20\tau) \\ 10e^{\tau}\cos(20\tau) \end{bmatrix}}_{\boldsymbol{m}(\tau)} \otimes I_2,\ \widehat{\mathsf{M}} = \underbrace{\begin{bmatrix} 0 & 0 & 0 \\ 0 & 1 & 20 \\ 0 & -20 & 1 \end{bmatrix}}_{\mathsf{M}} \otimes I_2 \tag{72}$$

in accordance to the definition in Assumption 1 with $\rho = 3,\ \ n = m = q = 2$.

By further using (72), the corresponding coefficient matrices $A_3, C_3$:

$$A_3 = 0.1\begin{bmatrix} -4 & 10 & -0.1 & 0.2 & 0.3 & 0.2 \\ -10 & 4 & 0.01 & 0.3 & -0.2 & 0.4 \end{bmatrix},\quad C_3 = 0.1\begin{bmatrix} 0 & 1 & 0.2 & 0 & 0 & 0.1 \\ 0 & -2 & 0.1 & 0.3 & -0.1 & 0 \end{bmatrix} \tag{73}$$

can be obtained, satisfying $\widetilde{A}_3(\tau) = A_3 M(\tau)$ and $\widetilde{C}_3(\tau) = C_3 M(\tau)$.

**Remark 11.** The matrices $A_3$ and $C_3$ can be determined via direct observations given that the structure of $M(\tau)$ in (72) is well ordered. On the other hand, considering $10e^{\tau}\sin(20\tau) = \theta_1$, $10e^{\tau}\cos(20\tau) = \theta_2$ in (71) yields polynomial equations $\widetilde{A}_3 = A_3 M(\theta_1, \theta_2)$, $\widetilde{C}_3 = A_3 M(\theta_1, \theta_2)$ in terms of $\theta_1, \theta_2$, which can be easily solved numerically via linear equations.

Apply Theorem 1 with (72) and (73) and assuming parameter $\eta_1 = 1$, $\eta_2 = \{\epsilon_i\}_{i=1}^{6} = 0$. Then it shows that the system (1) with system parameters (71) is stabilized by the controller

$$\boldsymbol{u}(t) = \begin{bmatrix} 1.7839 & -6.3792 \end{bmatrix} \boldsymbol{x}(t). \tag{74}$$

with $\gamma = 0.3468$.

To verify the synthesis result, we again apply the spectrum method in [2] to the resulting closed loop system. It yields $-0.1606 < 0$ as the real part of the rightmost characteristic root pair, which proves that the resulting closed loop system is stable.

It is important to mention that if the tuning factors $\eta_1$, $\eta_2$, $\{\epsilon_i\}_{i=1}^{\rho}$ in matrix $\check{\boldsymbol{\Theta}}$ are not given, then $\check{\boldsymbol{\Theta}} \prec 0$ in (22) becomes bilinear which can be solved by nonlinear solvers such as Penlab [8]. In addition, it should be stressed that the main motivation for applying a nonlinear solver is to find out the potential optimal values of tuning factors $\eta_1$, $\eta_2$, $\{\epsilon_i\}_{i=1}^{\rho}$ so they can be substituted into $\check{\boldsymbol{\Theta}} \prec 0$ to produce convex constraints.

*4.2.2. Synthesis considering both dissipativity and uncertainties*

Now consider the uncertainties defined in (54) and (56) with the same system in (71). For the sake of having a compact synthesis condition, we assume that $\Delta_i \in \mathbb{R}^{n\times n}$ and $\Lambda_i = F_i = \mathbf{O}_n, \forall i = 1 \cdots 10$. In addition, consider

$$H_1 = \begin{bmatrix} -0.1 & -0.7 \\ -0.3 & 0.3 \end{bmatrix},\ H_2 = \begin{bmatrix} 0.1 \\ -0.3 \end{bmatrix},\ H_3 = \begin{bmatrix} 0.1 & 0.2 \\ 0.1 & 0.1 \end{bmatrix},\ H_4 = \begin{bmatrix} 0.2 & 0.1 & -0.1 & 0.3 & 0.12 & -0.2 \\ 0.14 & 0.25 & 0.19 & -0.11 & -0.1 & -0.23 \end{bmatrix}$$

$$H_5 = \begin{bmatrix} 0.2 & 0.2 \\ 0.21 & 0.21 \end{bmatrix},\ H_6 = \begin{bmatrix} -0.12 & -0.14 \\ 0.01 & 0.2 \end{bmatrix},\ H_7 = \begin{bmatrix} 0.4 \\ -0.2 \end{bmatrix},\ H_8 = \begin{bmatrix} 0.12 & -0.14 \\ 0.01 & 0.2 \end{bmatrix},$$

$$H_9 = \begin{bmatrix} 0.2 & 0.1 & -0.1 & -0.14 & 0.1 & -0.1 \\ -0.1 & 0.3 & 0.2 & -0.1 & -0.1 & -0.1 \end{bmatrix},\ H_{10} = \begin{bmatrix} 0.22 & 0.23 \\ 0.22 & 0.23 \end{bmatrix},$$

$$G_i = \begin{bmatrix} 0.04 & 0.04 \\ 0.11 & 0.11 \end{bmatrix}, \forall i = 1 \cdots 5,\ G_i = \begin{bmatrix} 0.17 & 0.17 \\ 0.14 & 0.14 \end{bmatrix}, \forall i = 6 \cdots 10,\ \Xi_1 = \begin{bmatrix} 2.3 & 1 \\ * & 2.4 \end{bmatrix}, \Xi_2 = \begin{bmatrix} 1.5 & -0.5 \\ * & 2.9 \end{bmatrix},$$

$$\Xi_3 = \begin{bmatrix} 1.7 & 0.48 \\ * & 1.6 \end{bmatrix},\ \Xi_4 = \begin{bmatrix} 2.5 & 0.51 \\ * & 2 \end{bmatrix},\ \Xi_5 = \begin{bmatrix} 1.7 & 0.44 \\ * & 1.7 \end{bmatrix},\ \Xi_6 = \begin{bmatrix} 1.6 & 0.15 \\ * & 1.4 \end{bmatrix},\ \Xi_7 = \begin{bmatrix} 3.37 & -1.1 \\ * & 1.8 \end{bmatrix}$$

$$\Xi_8 = \begin{bmatrix} 1.54 & 0.13 \\ * & 1.34 \end{bmatrix},\ \Xi_9 = \begin{bmatrix} 2.7 & -0.65 \\ * & 1.87 \end{bmatrix},\ \Xi_{10} = \begin{bmatrix} 1.7 & 0.44 \\ * & 1.7 \end{bmatrix},\ \Gamma_i = \begin{bmatrix} -1.75 & -0.58 \\ * & -1.75 \end{bmatrix}, \forall i = 1 \cdots 10.$$

as the parameters for the uncertainties in (54) and (56).

By utilizing Theorem 2 with (72) and taking into account all the uncertainties, it follows that the uncertain system is robustly stabilized by the controller $\boldsymbol{u}(t) = \begin{bmatrix} 3.2847 & -16.7739 \end{bmatrix} \boldsymbol{x}(t)$ within theset $\mathcal{D}$ defined in (56) with the aforementioned parameters such that $\gamma = 0.62$.

Apply the spectrum method again to the resulting closed loop system without considering the uncertainties factors. It produces $-0.1773 < 0$ as thereal part of the rightmost characteristic root pair. Hence the resulting closed loop system in this case is stable.

**Remark 12.** In (72) the vector value function $\boldsymbol{m}(\tau)$ has been scaled compared to the form of (70). This is due to the fact that in some situations the differences among the eigenvalues of $\mathsf{F}$ are too large to be handled by the software. However, this should not be considered as a issue since mathematically a scaled $\boldsymbol{m}(\tau)$ does not introduce any conservatism or relaxation into the original synthesis conditions, rather it is a numerical implementation problem which can be easily circumvented.

## 5. Conclusion

In this paper a general synthesis method for linear DDS considering nonlinear distributed delay kernels has been presented. By constructing a Liapunov-Krasovskiĭ functional, sufficient conditions for the existence of a controller which stabilizes the closed loop system and ensures both dissipativity and robustness are derived in terms of LMIs. With the new proposed integral inequality, the synthesis conditions in this paper can produce non-conservative results with fewer decision variables compared to existing literatures. To show the validity and effectiveness of the proposed methodology, numerical examples have been investigated which can be efficiently solved via semidefinite programming algorithms.

## Acknowledgement

We gratefully thank the associate editor and anonymous reviewers for their constructive criticisms which have significantly improved the quality of this paper. The first author thanks his colleagues in The University of Auckland for their help with numerical simulations. Many thanks to Prof. Dimitri Breda and Prof. Johan Löfberg regarding the details of their Matlab toolboxes. Finally, the first author would like to express his sincere gratitude to Dr Bernard Guillemin whose help have greatly improved the language quality of this paper.

## Appendix A. Proof of Lemma 5

*Proof.* The proof is inspired by the results in [32, 35]. To begin with, we can conclude that the matrix $\mathsf{F}$ is well defined based on the results in Lemma 3.

Let $\boldsymbol{y}(\cdot) \in \mathscr{L}^2(\mathcal{K};\mathbb{R}^n_{\|2})$ such that

$$\boldsymbol{y}(\tau) := \boldsymbol{x}(\tau) - M^\top(\tau)(\mathsf{F}\otimes I_n)\int_{\mathcal{K}} M(\theta)\boldsymbol{x}(\theta)\mathsf{d}\theta. \tag{A.1}$$

Substituting (A.1) to the integral term $\int_{\mathcal{K}} \boldsymbol{y}^\top(\tau)U\boldsymbol{y}(\tau)\mathsf{d}\tau$ yields

$$\begin{aligned}\int_{\mathcal{K}} \boldsymbol{y}^\top(\tau)U\boldsymbol{y}(\tau)\mathsf{d}\tau = \int_{\mathcal{K}} \boldsymbol{x}^\top(\tau)U\boldsymbol{x}(\tau)\mathsf{d}\tau + \mathbf{z}^\top \int_{\mathcal{K}} (\mathsf{F}\otimes I_n)^\top M(\tau)UM^\top(\tau)(\mathsf{F}\otimes I_n)\mathsf{d}\tau\mathbf{z} \\ - 2\int_{\mathcal{K}} \boldsymbol{x}^\top(\tau)UM^\top(\tau)\mathsf{d}\tau(\mathsf{F}\otimes I_n)\mathbf{z}\end{aligned} \tag{A.2}$$

where $\mathbf{z} := \int_{\mathcal{K}} M(\theta)\boldsymbol{x}(\theta)\mathsf{d}\theta$.

Now apply the equality (8) to the term $M(\tau)U$ and consider the fact that $M(\tau) = \boldsymbol{m}(\tau)\otimes I_n$, we have

$$(\boldsymbol{m}(\tau)\otimes I_n)\,U = (I_d\otimes U)\,(\boldsymbol{m}(\tau)\otimes I_n)\,. \tag{A.3}$$

Moreover, since $U$ is a symmetric matrix, it also implies that

$$UM^\top(\tau) = U^\top M^\top(\tau) = \left(\boldsymbol{m}^\top(\tau)\otimes I_n\right)(I_d\otimes U), \tag{A.4}$$

given $(X\otimes Y)^\top = X^\top \otimes Y^\top$ holds.

Applying (A.4) to the term in (A.2) yields

$$\int_{\mathcal{K}} \boldsymbol{x}^\top(\tau) U M^\top(\tau) \mathrm{d}\tau (\mathsf{F} \otimes I_n)\mathbf{z} = \mathbf{z}^\top (I_d \otimes U)(\mathsf{F} \otimes I_n)\mathbf{z} = \mathbf{z}^\top(\mathsf{F} \otimes U)\mathbf{z}. \tag{A.5}$$

Furthermore, by (A.3) and the fact that $\mathsf{F} = \mathsf{F}^\top$, the term

$$\int_{\mathcal{K}} (\mathsf{F} \otimes I_n)^\top M(\tau) U M^\top(\tau)(\mathsf{F} \otimes I_n)\mathrm{d}\tau$$

in (A.2) can be reformulated into

$$\int_{\mathcal{K}} (\mathsf{F} \otimes I_n)(I_d \otimes U) M(\tau) M^\top(\tau)(\mathsf{F} \otimes I_n)\mathrm{d}\tau. \tag{A.6}$$

Now, utilizing $M(\tau) = \boldsymbol{m}(\tau) \otimes I_n$ produces

$$\begin{aligned}\int_{\mathcal{K}} (\boldsymbol{m}(\tau) \otimes I_n)\left(\boldsymbol{m}^\top(\tau) \otimes I_n\right)(\mathsf{F} \otimes I_n)\mathrm{d}\tau &= \left(\int_{\mathcal{K}} \boldsymbol{m}(\tau)\boldsymbol{m}^\top(\tau)\mathrm{d}\tau\right) \otimes I_n(\mathsf{F} \otimes I_n) \\ &= \left[\mathsf{F}^{-1} \otimes I_n\right](\mathsf{F} \otimes I_n) = I_{nd}. \end{aligned} \tag{A.7}$$

As a result, (A.6) can be simplified into a single matrix $(\mathsf{F} \otimes I_n)(I_d \otimes U) = \mathsf{F} \otimes U$. Substituting this term into the original expression (A.2) and also considering the relation in (A.5) produces

$$\int_{\mathcal{K}} \boldsymbol{y}^\top(\tau) U \boldsymbol{y}(\tau)\mathrm{d}\tau = \int_{\mathcal{K}} \boldsymbol{x}^\top(\tau) U \boldsymbol{x}(\tau)\mathrm{d}\tau - [*](\mathsf{F} \otimes U)\left[\int_{\mathcal{K}} M(\tau)\boldsymbol{x}(\tau)\mathrm{d}\tau\right]. \tag{A.8}$$

Given $U \succeq 0$, it concludes the validity of (16). This finishes the proof. ∎

## Appendix B. Proof of Lemma 6

*Proof.* The proof is inspired by the strategies illustrated in [4]. Consider $\mathcal{F} = \mathcal{D}$. We first need to find an equivalent condition for the well-posedness of $(I_m - \Delta F)^{-1}$ for all $\Delta \in \mathcal{D}$, where $\mathcal{D}$ has been defined in (58). It is obvious that $(I_m - \Delta F)^{-1}$ is well defined for all $\Delta \in \mathcal{D}$ if and only if $\forall \Delta \in \mathcal{D}, (I_m - \Delta F) \in \mathbb{R}^{m\times m}_{[m]}$. Furthermore, we know that $\forall \Delta \in \mathcal{D}$, $(I_m - \Delta F) \in \mathbb{R}^{m\times m}_{[m]}$ if and only if $\forall \Delta \in \mathcal{D}$, $(I_m - \Delta F)^\top (I_m - \Delta F) \succ 0$ according to the property of quadratic forms.

Let $\mathbb{R}^m \ni \boldsymbol{\mu} := \Delta F \boldsymbol{\theta}$ and $\mathbb{R}^p \ni \boldsymbol{\omega} := F\boldsymbol{\theta}$ with $\boldsymbol{\theta} \in \mathbb{R}^m$, we can conclude that $\forall \Delta \in \mathcal{D}$, $(I_m - \Delta F)^\top (I_m - \Delta F) \succ 0$ if and only if $\forall \boldsymbol{\theta} \in \mathbb{R}^m \backslash \mathbf{0}$, $\forall \Delta \in \mathcal{D}$, $\boldsymbol{\theta}^\top (I_m - \Delta F)^\top (I_m - \Delta F)\boldsymbol{\theta} > 0$ which is equivalent to

$$\begin{bmatrix}\boldsymbol{\theta} \\ \boldsymbol{\mu}\end{bmatrix}^\top \begin{bmatrix} I_m & -I_m \\ * & I_m \end{bmatrix} \begin{bmatrix}\boldsymbol{\theta} \\ \boldsymbol{\mu}\end{bmatrix} > 0, \ \forall \begin{bmatrix}\boldsymbol{\theta} \\ \boldsymbol{\mu}\end{bmatrix} \in \mathcal{M}, \tag{B.1}$$

$$\text{with } \mathcal{M} \in \left\{ \begin{bmatrix}\boldsymbol{\theta} \\ \boldsymbol{\mu}\end{bmatrix} \in \mathbb{R}^{2m}_{\backslash\{\mathbf{0}\}} \;\middle|\; [*] \begin{bmatrix} \Theta_1^{-1} & \Theta_2 \\ * & \Theta_3 \end{bmatrix} \underbrace{\begin{bmatrix} F & \mathsf{O}_{p\times m} \\ \mathsf{O}_m & I_m \end{bmatrix} \begin{bmatrix}\boldsymbol{\theta} \\ \boldsymbol{\mu}\end{bmatrix}}_{\mathbf{col}\left[I_p\boldsymbol{\omega},\ \Delta\boldsymbol{\omega}\right]} \geq 0 \right\},$$

by considering $\boldsymbol{\mu} := \Delta F \boldsymbol{\theta}$ and $\boldsymbol{\mu} = \Delta \boldsymbol{\omega}$.

Invoking S-procedure to (B.1) produces the conclusion that (B.1) is true if and only if

$$\exists \alpha > 0 : \begin{bmatrix} I_m & -I_m - \alpha F^\top \Theta_2 \\ * & I_m - \alpha\Theta_3 \end{bmatrix} - \begin{bmatrix} \alpha F^\top \\ \mathsf{O}_{m\times p} \end{bmatrix} (\alpha\Theta_1)^{-1} \begin{bmatrix} \alpha F & \mathsf{O}_{p\times m} \end{bmatrix} \succ 0. \tag{B.2}$$

By applying Schur complement to (B.2) with the fact that $\Theta_1^{-1} \succ 0$, it produces (57). Furthermore, if $\mathcal{F} \subset \mathcal{D}$, then the well posed condition is valid under $\mathcal{F}$ as well.

As a result, $(I_m - \Delta F)^{-1}$ is well defined for all $\Delta \in \mathcal{D}$ if and only if (57) holds.

Now we shall proceed to prove the rest of the conclusions in Lemma 6. By the definition of positive definite matrices, we know that $\forall \Delta \in \mathcal{D}$, $\Phi + \mathbf{Sy}\left[G(I_m - \Delta F)^{-1}\Delta H\right] \prec 0$ if and only if $\forall \Delta \in \mathcal{D}$, $\forall \mathbf{x} \in \mathbb{R}^n$, $\mathbf{x}^\top \left(\Phi + \mathbf{Sy}\left[G(I_m - \Delta F)^{-1}\Delta H\right]\right)\mathbf{x} < 0$. Letting $\mathbb{R}^m \ni \boldsymbol{\varrho} := (I_m - \Delta F)^{-1}\Delta H\mathbf{x}$ and multiplying terms $(I_m - \Delta F)$ on the two sides of the equation yields $\boldsymbol{\varrho} = \Delta H\mathbf{x} + \Delta F\boldsymbol{\varrho}$. By further defining $\mathbb{R}^p \ni \mathbf{z} := H\mathbf{x} + F\boldsymbol{\varrho}$, the aforementioned relations can be simplified into a compact equation $\boldsymbol{\varrho} = \Delta\mathbf{z}$. By considering intermediate variables $\mathbf{z}$ and $\boldsymbol{\varrho}$ with the relations $\boldsymbol{\varrho} = \Delta\mathbf{z}$ and $\mathbf{z} := H\mathbf{x} + F\boldsymbol{\varrho}$, we can reformulate the original proposition into

$$\begin{bmatrix}\mathbf{x}\\ \boldsymbol{\varrho}\end{bmatrix}^\top \begin{bmatrix}\Phi & G\\ * & 0_m\end{bmatrix}\begin{bmatrix}\mathbf{x}\\ \boldsymbol{\varrho}\end{bmatrix} < 0,\ \forall \begin{bmatrix}\mathbf{x}\\ \boldsymbol{\varrho}\end{bmatrix} \in \mathcal{X} \tag{B.3}$$

$$\text{with } \mathcal{X} \in \left\{ \begin{bmatrix}\mathbf{x}\\ \boldsymbol{\varrho}\end{bmatrix} \in \mathbb{R}^{n+m}_{\setminus\{\mathbf{0}\}} \;\middle|\; \begin{bmatrix}*\end{bmatrix}\begin{bmatrix}\Theta_1^{-1} & \Theta_2\\ * & \Theta_3\end{bmatrix}\underbrace{\begin{bmatrix}H & F\\ 0_{m\times p} & I_m\end{bmatrix}\begin{bmatrix}\mathbf{x}\\ \boldsymbol{\varrho}\end{bmatrix}}_{\mathbf{col}\left[I_p\mathbf{z},\ \ \Delta\mathbf{z}\right]} \geq 0 \right\},$$

which is equivalent to (58).

Clearly, the structure of (B.3) enables us to implement S-procedure so that we can conclude (B.3) is true if and only if

$$\exists \kappa > 0 : \begin{bmatrix}\Phi & G + \kappa H^\top\Theta_2\\ * & \kappa F^\top\Theta_2 + \kappa\Theta_2^\top F + \kappa\Theta_3\end{bmatrix} + \begin{bmatrix}\kappa H^\top\\ \kappa F^\top\end{bmatrix}(\kappa\Theta_1)^{-1}\begin{bmatrix}\kappa H & \kappa F\end{bmatrix} \prec 0. \tag{B.4}$$

It should be emphasized that for the cases of having a single constraint like (B.3), S-procedure is able to produce an equivalent condition. Invoking Schur complement on equation (B.4) and considering $\Theta_1^{-1} \succ 0$ produces (59) which prove the equivalence between (59) and (58). Finally, if $\mathcal{F} \subset \mathcal{D}$, then (59) becomes a sufficient condition for (58).

For the situation when $\Theta_1^{-1} = 0_p$, (B.4) and (B.2) become (61) and (60), respectively. As we have mentioned in dealing with the supply function (4), there is no need to consider $\Theta_1$. This is because both expressions in this case does not require Schur complement to derive a single inequality.

Since all the aforementioned derivations can be conducted inversely under the same assumptions, therefore it proves the validity of Lemma 6. Finally, it is important to mention that (57) and (59) are able to handle the uncertainty structure $G\Delta(I_p - F\Delta)^{-1}H$, which can be comprehended by considering the equalities $\det(I_p - F\Delta) = \det(I_m - \Delta F)$ and $\Delta(I_p - F\Delta)^{-1} = (I_m - \Delta F)^{-1}\Delta$. ■